\documentclass[preprint]{aastex}
\usepackage{amssymb}
\usepackage{stmaryrd}
\usepackage{amssymb,amsmath, epsfig, epsf, lscape}

\def\beq{\begin{equation}}
\def\eeq{\end{equation}}
\def\bey{\begin{eqnarray}}
\def\eey{\end{eqnarray}}

\def\mpc{\, h^{-1}{\rm {Mpc}}}

\def\mpci{\, h{\rm {Mpc}}^{-1}}

\def\msun{\, h^{-1}{\rm M_\odot}}

\def\gs{\mathrel{\raise1.16pt\hbox{$>$}\kern-7.0pt
\lower3.06pt\hbox{{$\scriptstyle \sim$}}}}
\def\ls{\mathrel{\raise1.16pt\hbox{$<$}\kern-7.0pt
\lower3.06pt\hbox{{$\scriptstyle \sim$}}}}
\def\gtsima{$\; \buildrel > \over \sim \;$}
\def\ltsima{$\; \buildrel < \over \sim \;$}
\def\prosima{$\; \buildrel \propto \over \sim \;$}
\def\gsim{\lower.5ex\hbox{\gtsima}}
\def\lsim{\lower.5ex\hbox{\ltsima}}
\def\simgt{\lower.5ex\hbox{\gtsima}}
\def\simlt{\lower.5ex\hbox{\ltsima}}
\def\simpr{\lower.5ex\hbox{\prosima}}

\slugcomment{To appear in {\it The Astrophysical Journal}}
\shorttitle{Exploring the Local Universe with reConstructed Initial Density field} \shortauthors{Wang H.Y. et al.}
\received{2014}

\begin{document}
\title {ELUCID - Exploring the Local Universe with reConstructed Initial Density field
I: Hamiltonian Markov Chain Monte Carlo Method with Particle Mesh Dynamics}
\author{Huiyuan Wang\altaffilmark{1,2}, H.J. Mo\altaffilmark{1,2}, Xiaohu Yang\altaffilmark{3,4}, Y. P. Jing\altaffilmark{4} and W. P. Lin\altaffilmark{3}}

\altaffiltext{1}{Key Laboratory for Research in Galaxies and
Cosmology, Department of Astronomy, University of Science and
Technology of China, Hefei, Anhui 230026, China;
whywang@mail.ustc.edu.cn} \altaffiltext{2}{Department of
Astronomy, University of Massachusetts, Amherst MA 01003-9305,
USA} \altaffiltext{3}{Key Laboratory for Research in Galaxies and
Cosmology, Shanghai Astronomical Observatory, Shanghai 200030,
China} \altaffiltext{4}{Center for Astronomy and Astrophysics,
Shanghai Jiao Tong University, Shanghai 200240, China}

\begin{abstract}
Simulating the evolution of the local universe is important for studying
galaxies and the intergalactic medium in a way free of cosmic variance. Here we present a method to reconstruct the initial linear density
field from an input non-linear density field, employing the Hamiltonian Markov Chain Monte Carlo (HMC) algorithm
combined with Particle Mesh (PM) dynamics. The HMC+PM method is applied to cosmological simulations, and
the reconstructed linear density fields are then evolved to the present day with $N$-body simulations.
The constrained simulations so obtained accurately reproduce both the amplitudes and
phases of the input simulations at various $z$. Using a PM model with
a grid cell size of $0.75\mpc$ and 40 time-steps in the HMC can recover more than half of the phase information down to a scale $k\sim0.85\mpci$ at high $z$ and to $k\sim3.4\mpci$ at $z=0$, which represents a significant improvement over similar reconstruction models in the literature, and indicates that our model can reconstruct
the formation histories of cosmic structures over a large dynamical range. Adopting PM models with higher
spatial and temporal resolutions yields even better reconstructions, suggesting that our method is limited more by the availability of computer resource than by principle. Dynamic models of structure evolution
adopted in many earlier investigations can induce non-Gaussianity in the reconstructed linear density
field, which in turn can cause large systematic deviations in the predicted halo mass function. Such deviations
are greatly reduced or absent in our reconstruction.
\end{abstract}


\section{Introduction}
\label{sec_intro}

A key step in understanding the physical processes of galaxy formation
is to investigate the correlations and interactions among galaxies,
baryonic gas [particularly interstellar media and
intergalactic media (IGM)] and dark matter. Many observational
programs have been carried out for galaxies and the gas components.
Large redshift surveys, such as the Sloan Digital Sky Survey \citep[SDSS;][]{York_etal00},
can now provide a huge amount of information
about the intrinsic properties of galaxies and their clustering in space.
Interstellar gas closely associated with galaxies can be observed
through its 21cm emission of neutral hydrogen gas
\citep[e.g.][]{Koribalski_etal04,Springob_etal05,Giovanelli_etal05},
and through millimeter/submillimeter emissions of molecular gas
\citep[e.g.][]{Young_etal95,Saintonge_etal11}. The diffuse cold/warm components in the IGM can be
studied using quasar absorption line systems \citep[e.g.][]{Savage_etal98,Fang_etal02,Werk_etal13},
while the hot components can be studied through X-ray observations
and through their Sunyaev-Zel'dovich effect in the
Cosmic Microwave Background \citep[see][for a review]{Carlstrom_etal02}.

In order to make full use of the potential of these observational
data, one has to develop optimal strategies. Clearly, it is important
to have as much constraints on the dark matter component.
As the dominating mass component of the cosmic density
field, the distribution of dark matter relative to the galaxy population
and the IGM can provide important information about the
large-scale environments within which galaxies and the gas
components evolve. In particular, any constraints on the
evolutionary histories of the dark matter structures within
which the observed galaxies and gas reside can provide direct
information on how the interactions between the dark matter
and baryonic components shape the observed properties of
galaxies and IGM.

Unfortunately, the details about the distribution and evolution
of the dark matter component in the observed universe is not
directly available; current
observations of gravitational lensing can only provide constraints
on the properties of the mass density field in a statistical
way but not on an object-to-object basis. One promising way
to make progress in this direction is to reconstruct the initial
(linear) density field from which the observed structures in
the present-day universe form. This is now possible owing to the well established paradigm, the $\Lambda$CDM
model, within which the relationship between galaxy systems
(individual galaxies, groups and clusters of galaxies)
and the underlying dark matter distribution can be modeled
quite accurately through the connection between the distribution of dark
matter halos and the overall mass distribution.  With such reconstructed
initial conditions, one can use high-resolution numerical
simulations  (usually referred to as constrained simulations) to
reproduce the time evolution of the density field in the local Universe.
This can not only be used to trace the present-day
environments in which observed galaxies reside,  but also
provide the formation histories of individual structures to
understand the interaction between dark matter and baryonic
components. Indeed, together with galaxy formation and baryonic
physics, the reconstructed initial conditions allow one to
simulate the formation and evolution of the galaxy
population and IGM in the local Universe. Such an approach
is particularly powerful in constraining theory with observation,
because many observations, such as those for dwarf galaxies
and for low-$z$ quasar absorption line systems, can only
be carried out in relatively small volumes, and so the cosmic variance
is usually a big issue.  The uncertainties due to the cosmic variance
can be minimized by making comparisons between
observations and model predictions for systems that have
both the same environments and the same formation histories,
which is exactly an accurate constrained simulation is capable of
doing.

Numerous attempts have been made to develop methods to reconstruct
the initial conditions of structure formation in the local universe
using galaxy distributions and/or peculiar velocities.
\citet{HoffmanRibak91} developed a method to construct
Gaussian random fields that are subjected to various constraints
\citep[see also][]{Bertschinger87,vandeWeygaert96,Klypin_etal03,KitauraEnblin08}. \cite{Klypin_etal03}
improved this method by using Wiener Filter \citep[see e.g.][]{Zaroubi_etal95} to deal
with sparse and noisy data.
Gaussian density fields constrained by the peculiar velocities of galaxies
in the local universe have been used to set up the initial conditions
for constrained simulations \citep[e.g.][]{Kravtsov_etal02,Klypin_etal03,Gottloeber_etal10}.
Note, however, that the basic underlying assumption in this method is
that the linear theory is valid on all scales \citep{Klypin_etal03,Doumler_etal13}.

\citet{NusserDekel92} proposed a method which employs
quasi-linear dynamics for structure evolution. This method traces the
non-linear mass (galaxy) density field back in time to the linear
regime according to the Zel'dovich approximation in
Eulerian space \citep[see][for another related approach]{Peebles89}.
Under the assumption of the absence of multi-streaming
(shell-crossing), \citet{Brenier_etal03}
found that the reconstruction problem can be treated as an
instance of optimal mass transportation, and developed a
Monge-Amp\'ere-Kantorovich (MKA) method to recover the
particle displacement field \citep[see also][]{Frisch_etal02}. These two methods are valid only
on scales where a one-to-one relation between the Lagrangian
and Eulerian coordinates exists. Furthermore the two methods
did not take account of any priors about the statistical properties of
the initial density field, and so the reconstructed initial density field
is not guaranteed to be Gaussian \citep{Kolatt_etal96,Doumler_etal13}.

In order to achieve a high reconstruction precision and simultaneously
to avoid non-Gaussianity, several hybrid approaches have been
proposed.  For example, \cite{Lavaux10} used the MKA method to
generate constraints from observations, requiring
the constraints to have Gaussian distributions. \cite{Doumler_etal13} extended
the method by adding an inverse Zel'dovich approximation.

More recently, Bayesian approaches have been proposed,
in which the initial (linear) density field is sampled from a
posterior probability distribution function consisting of a
Gaussian prior and a likelihood \citep{JascheWandelt13,Kitaura13,Wang_etal13,Heb_etal13}.
In such an approach, a specific dynamic model of structure evolution has to be
adopted to link the linear density field to the observed galaxy
distribution \citep[e.g.][]{JascheWandelt13,Kitaura13} or
to the present-day mass density field inferred from other
means \citep[hereafter W13]{Wang_etal13}. The models used in the literature include
the modified Zel'dovich approximation \citep[MZA,][hereafter TZ12]{TZ12}, the second order
Lagrangian perturbation theory (2LPT), and the augmented
Lagrangian perturbation theory \citep[ALPT,][]{KitauraHb13}. In such a Bayesian approach,
different initial density fields are evolved forward to predict the non-linear density field today, and their merits
are evaluated in terms of adopted priors and how well their
predicted non-linear density fields match the constraining input field \citep{JascheWandelt13}.
The advantage of such an approach is twofold. First, it automatically
takes into account priors for the initial density field. Second,
as a `forward' approach, it is not limited by the development
of multi-stream flows, as long as the adopted
dynamical model can follow them accurately.

Clearly, the performance of the Bayesian approach depends on
the accuracy of the dynamical model adopted to follow the structure
evolution. As to be detailed in Section \ref{sec_dsf}, the dynamical
models so far adopted work only accurately at wave-numbers
$k\simlt0.5\mpci$, i.e. only in the quasi-nonlinear regime.
They can not properly account for the evolution of highly
non-linear structures, such as massive clusters, filaments and sheets,
where shell-crossing is frequent. If such an approximate model
is adopted to reconstruct the structures in the non-linear regime,
bias can be introduced into the reconstructed initial conditions.
Such bias sometimes can be very significant and may greatly reduce
the usefulness of the reconstructed initial density field.

In this first paper of a series, we develop a method combining
the Bayesian reconstruction approach with a much more
accurate dynamic model of structure evolution,  the
Particle Mesh (PM) model. The PM technique has been
commonly adopted in $N$-body codes to evaluate
gravitational forces on relatively large scales \citep[see e.g.][]{White_etal83,KlypinShandarin83,JingSuto02,Springel05},
and can follow the structure evolution accurately as long as
the grid cells and time steps are chosen sufficiently small.
We show that our method is limited more by the availability of computer
resource than by principle. Even with a modest computer
resource, the reconstruction accuracy our method can achieve is
already much higher than those of the other methods
in the literature.

The structure of the paper is organized as follows.
In Section \ref{sec_method}, we describe our reconstruction
method based on Hamiltonian Markov Chain Monte Carlo
(hereafter HMC). Section \ref{sec_sim} describes the $N$-body
simulations used for testing our method, and how the quality
of the PM model depends on model parameters.
In Section \ref{sec_app}, we test our HMC$+$PM method by applying
it to high-resolution numerical simulations. In Section \ref{sec_cs}, we
use $N$-body simulations to follow the structure evolution
seeded by the reconstructed linear density field and compare
the results with the original simulations to examine various aspect
of our method. Section \ref{sec_dsf} shows the comparison between
the results of HMC reconstructions adopting different dynamical
models. Finally, in Section \ref{sec_sum} we summarize our main
results and make some further discussions. Many abbreviations, terms
and quantities are used in the text. To avoid confusion,
we provide a list of them together with their definitions
in Appendix \ref{app_abb}.

\section{The reconstruction method}
\label{sec_method}

Our objective is to reconstruct the linear (or initial) density field
from an input non-linear density field, $\rho_{\rm inp}({\bf x})$, at low
redshift, such as at the present day. The input density field
can either be that from an accurate cosmological simulation
($\rho_{\rm si}$), as is the case here for testing the method of
reconstruction, or that reconstructed from observations, such as a
galaxy redshift survey that can be used to trace the current mass
density field in the Universe. There are two constraints on
this linear density field, which can be specified by its Fourier
transform, $\delta({\bf k})$. First, the linear density field is
required to be consistent with a chosen cosmology. We assume
the standard $\Lambda$CDM model, and so the linear density field
obeys a Gaussian distribution with variance given by the linear
power spectrum, $P_{\rm lin}(k)$. Second, the modeled density field,
$\rho_{\rm mod}({\bf x})$, evolved from $\delta({\bf k})$ according to
a chosen dynamical model of structure evolution, should match the input density
field, $\rho_{\rm inp}({\bf x})$, as close as possible.

Owing to the complexity of the problem, such as the very high
dimensionality of the parameter space, and
the uncertainty and incompleteness in the
constraining field, $\rho_{\rm inp}({\bf x})$, the solution may not be unique.
We therefore follow W13 and construct a
posterior probability distribution for $\delta({\bf k})$ given
$\rho_{\rm inp}({\bf x})$ as
\begin{eqnarray}
\emph{Q}(\delta_j({\bf k})|\rho_{\rm inp}({\bf x}))
&=&{\rm e}^{-\chi^2}\times G (\delta({\bf k}))\nonumber\\
&=&
{\rm e}^{-\sum_{\bf x}[\rho_{\rm mod}({\bf x})-\rho_{\rm inp}({\bf
x})]^2\omega({\bf x})/2\sigma_{\rm inp}^2({\bf x})}\times\prod_{\bf
k}^{\rm half}\prod_{j={\rm re}}^{\rm im}\frac{1}{[\pi P_{\rm lin}(k)]^{1/2}}{\rm
e}^{-[\delta_j({\bf k})]^2/P_{\rm lin}(k)}\label{eq_post}\,,
\end{eqnarray}
where $\sigma_{\rm inp}$ is the statistical uncertainties
in $\rho_{\rm inp}$, while $\omega({\bf x})$ is a weight
function used to account for possible incompleteness.
The subscripts $j = {\rm re},\, {\rm im}$ denote the real and
imaginary parts, respectively. Since $\delta({\bf k})$ is the Fourier transform
of a real field, we have $\delta({\bf k}) = \delta^{\ast}(-{\bf k})$ so
that only the Fourier modes in the upper half-space are needed. All
these fields are to be sampled in a periodic box of length $L$ on a
side, divided into $N_{\rm c}$ grid cells in each dimension.
The term, $G (\delta({\bf k}))$, in the equation represents
the first constraint mentioned above, i.e. the prior Gaussian
distribution of $\delta({\bf k})$, while the  ${\rm e}^{-\chi^2}$
term accounts for the second constraint and can be regarded
as the likelihood for $\rho_{\rm mod}({\bf x})$ given $\rho_{\rm inp}({\bf x})$.

Our purpose is thus to seek the solutions of $\delta({\bf k})$
that maximize the posterior probability distribution function
$\emph{Q}(\delta_j({\bf k})|\rho_{\rm inp}({\bf x}))$. As demonstrated
in W13, the Hamiltonian Markov Chain Monte Carlo
(HMC) technique developed by \cite{Duane_etal87} and \cite{Neal96}
can help us to achieve this goal, as it can sample a posterior
distribution in a large, multi-dimensional parameter space
very efficiently \citep{Hanson01}. This method has been widely applied
in astrophysics and cosmology \citep[e.g.][]{Hajian07,Taylor_etal08,JascheKitaura10,Kitaura_etal12,JascheWandelt13,Kitaura13,Wang_etal13,Heb_etal13}.
In particular, \cite{JascheWandelt13}, \cite{Kitaura13}, W13 and \cite{Heb_etal13} adopted this technique to
reconstruct the initial conditions for the local universe.
A brief description about the method is given in Section \ref{sec_HMCMC}.

In order to predict $\rho_{\rm mod}({\bf x})$
from $\delta({\bf k})$, a structure formation model is required
to evolve the cosmic density field. In this paper,
we first use the Zel'dovich approximation to generate the particle distribution at a given high redshift, and
then use the Particle-Mesh (PM) model to evolve the cosmic density
field traced by these particles. These techniques are well developed
and have been adopted in many cosmological simulations \citep{Zeldovich70,White_etal83,KlypinShandarin83,JingSuto02,Springel05}.
In Section \ref{sec_sfm}
we provide key equations used in our reconstruction.
Because the HMC method utilizes the gradients,
$\partial\rho_{\rm mod}({\bf x})/\partial\delta_{\rm j}({\bf k})$,
to suppress random walks in the MCMC implementation and
to improve the efficiency, it is necessary to obtain these gradients
from the adopted dynamic model for structure evolution. The derivation of
the Hamiltonian force, which is the combination of these gradients,
is described in Section \ref{sec_hf} and in Appendix \ref{app_hf}.

\subsection{The Hamiltonian Markov Chain Monte Carlo Algorithm}
\label{sec_HMCMC}

The algorithm is designed to sample the posterior distribution (our
target distribution) in a way analogous to solving a physical system
in Hamiltonian dynamics. The `potential' of the system is defined
as the negative of the natural logarithm of the target distribution,
\begin{equation}\label{targetprob}
\psi[\delta_j({\bf k})]\equiv -\ln (Q)
=\sum_{\bf k}^{\rm half}\ln[\pi P_{\rm lin}(k)]+\sum_{\bf
k}^{\rm half}\sum_{j={\rm re}}^{\rm im}\frac{[\delta_j({\bf k})]^2}{P_{\rm
lin}(k)}+\sum_{\bf x}\frac{[\rho_{\rm mod}({\bf x})
-\rho_{\rm inp}({\bf x})]^2\omega({\bf x})}{2\sigma_{\rm inp}^2({\bf x})}\,.
\end{equation}
For each $\delta_j({\bf k})$, a momentum variable, $p_j({\bf k})$,
is introduced and the Hamiltonian of the system is constructed as
\begin{equation}\label{eqhamilton}
H=\sum_{\bf k}^{\rm half}\sum_{j={\rm re}}^{\rm im}\frac{p_j^2({\bf
k})}{2m_j({\bf k})}+\psi[\delta_j({\bf k})]\,,
\end{equation}
where $m_j({\bf k})$ is a fictitious mass.

The statistical properties of the system is characterized by the partition
function, $\exp(-H)$, which is separated into a Gaussian
distribution in the momenta ${p_j({\bf k})}$ multiplied by the target
distribution:
\begin{equation}
\exp(-H)=\emph{Q}[\delta_j({\bf k})|\rho_{\rm p}({\bf x})]
\prod_{\bf k}^{\rm half}\prod_{j={\rm re}}^{\rm im}{\rm e}^{-\frac{p_j^2({\bf
k})}{2m_j({\bf k})}}\label{eq_eh}\,.
\end{equation}
As can be seen, the target distribution can be obtained by sampling
this partition function and then marginalizing over the momenta.

In order to sample the distribution, we first pick
a set of momenta $p_j({\bf k})$ randomly from the
multi-dimensional, un-correlated Gaussian distribution with
variances $m_j({\bf k})$. We then evolve the system from the
starting point, $[\delta_j({\bf k}), p_j({\bf k})]$, according to the
Hamilton equations. In practice, the leapfrog technique is
adopted to integrate the ``equations of motion'':
\begin{equation}\label{eq_lf1}
p_j({\bf k},t+\tau/2)=p_j({\bf k},t)-\frac{\tau}{2}\frac{\partial
H}{\partial \delta_j(\bf k)}{\Bigg\vert}_t\,;
\end{equation}
\begin{equation}\label{eq_lf2}
\delta_j({\bf k},t+\tau)=\delta_j({\bf k},t)+\frac{\tau}{m_j({\bf
k})}p_j({\bf k},t+\tau/2)\,;
\end{equation}
\begin{equation}\label{eq_lf3}
p_j({\bf k},t+\tau)=p_j({\bf
k},t+\tau/2)-\frac{\tau}{2}\frac{\partial H}{\partial
\delta_j(\bf k)}{\Bigg\vert}_{t+\tau}\,,
\end{equation}
where $\tau$ is the time increment for the leapfrog step. The
equations are integrated for $n$ leapfrog steps
(referred to as one chain step) to a final point,
$[\delta'_j({\bf k}),p'_j({\bf k})]$, in phase space. This final
state is accepted with a probability
\begin{equation}\label{paccept}
p={\rm min}\left \{1, {\rm e}^{-[H(\delta'_j({\bf k}), p'_j({\bf
k})) -H(\delta_j({\bf k}), p_j({\bf k}))]}\right \}\,.
\end{equation}
Since the Hamiltonian of a physical system is conserved, the
acceptance rate should in principle be unity, which is one of the
main advantages of the HMC method. In practice, however,
rejection can occur because of numerical errors.
The above process is repeated by randomly picking a new
set of momenta.

As one can see from equations (\ref{eq_lf1}) - (\ref{eq_lf3}), the Hamiltonian force,
$\partial H/\partial\delta_j(\bf k)$, is the most important
quantity to compute in order to evolve the system forward in
$t$. Combining equation (\ref{targetprob}) and (\ref{eqhamilton}),
we can write
\begin{equation}
\frac{\partial H}{\partial
\delta_j(\bf k)}=\frac{2\delta_j({\bf k})}{P_{\rm lin}(k)}
+\frac{\partial\chi^2}{\partial\delta_j({\bf k})}
=\frac{2\delta_j({\bf k})}{P_{\rm lin}(k)}+F_{j}({\bf k})\,,
\label{eq_dp}
\end{equation}
where
\begin{equation}
F_j({\bf k})\equiv \frac{\partial\chi^2}{\partial\delta_j({\bf k})}
\end{equation}
is the likelihood term of the Hamiltonian force to be discussed
in great detail in Section \ref{sec_hf} and in Appendix \ref{app_hf}. In order to proceed,
two other parameters, the Hamiltonian mass $m_j({\bf k})$
and the pseudo time $T=n\tau$, have to be specified.
As shown in \citet{Hanson01}, these two parameters have to be chosen carefully, because they can affect the
sampling efficiency significantly. To avoid resonant trajectory,
$T$ must be randomized. We thus randomly pick $n$ and $\tau$
from two uniform distributions in the range of $[1,n_{\rm max}]$
and $[0,\tau_{\rm max}]$, respectively. Following W13,
we set $n_{\rm max}=13$ and $\tau_{\rm max}\sim0.1$ and define
the Hamiltonian mass as,
\begin{equation}\label{eq_Hmass}
m_j({\bf k}) \equiv m(k) = \frac{2}{P_{\rm lin}(k)} +
\sqrt{\frac{\sum_{j={\rm re}}^{\rm im}\langle F^2_j({\bf k})\rangle_{\bf
k}}{P_{\rm lin}(k)}}\,,
\end{equation}
where $\langle\cdot\cdot\cdot\rangle _{\bf k}$ represents average
over the phase of ${\bf k}$. The suitability of this mass definition
is detailed in W13.

\subsection{Dynamic Model of Structure Evolution}
\label{sec_sfm}

In this section, we describe the dynamic model of structure evolution
adopted here to link the final density field, $\rho_{\rm mod}({\bf x})$, with
the initial linear density field, $\delta({\bf k})$. We first use the
Zel'dovich approximation to generate the positions and velocities of
particles at a given initial redshift $z_{\rm i}$  based on the
linear density field $\delta({\bf k})$. We then use the Particle-Mesh
(PM) technique to evolve the initial density field to the present day
to  obtain $\rho_{\rm mod}({\bf x})$.

\subsubsection{The Zel'dovich Approximation}
\label{sec_za}

For a particle with Lagrangian position ${\bf q}$, we use the Zel'dovich
approximation to derive its position ${\bf r}_{\rm i}({\bf q})$ and
velocity ${\bf v}_{\rm i}({\bf q})$ at initial redshift $z_{\rm i}$ as,
\begin{equation}\label{eq_za1}
{\bf r}_{\rm i}({\bf q})={\bf q}+{\bf s}({\bf q})\,;
\end{equation}
\begin{equation}\label{eq_za2}
{\bf v}_{\rm i}({\bf q})=H_{\rm i}a^2_{\rm i}f(\Omega_{\rm i}){\bf s}({\bf q})\,.
\end{equation}
Here, $H_{\rm i}$ is the Hubble constant at $z_{\rm i}$;
$a_{\rm i} =1/(1+z_{\rm i})$ is the scale factor;
and $f(\Omega)={\rm d}\ln D/{\rm d}\ln a$ with $D(a)$ the linear
growth factor. The Fourier transform of the displacement
field ${\bf s}({\bf q})$ is given by
\begin{equation}\label{eq_za3}
{\bf s}({\bf k})=\frac{i\bf k}{k^2}D(a_{\rm i})\delta({\bf k})\,.
\end{equation}
Note that the particle velocities ${\bf v}({\bf q})$
are not the peculiar velocities; the peculiar velocities
are ${\bf u}={\bf v}({\bf q})/a$. This definition of velocity allows
us to get rid of the $\dot{a}/a$ terms in the equations of motion
to be shown in the following. We use ${\bf x}$ to indicate the position of a
grid cell, and ${\bf q}$ and ${\bf r}$  to denote the Lagrangian
and Eulerian coordinates of a particle, respectively,
all in co-moving units. Both ${\bf x}$ and ${\bf q}$ are
regularly spaced, while ${\bf r}$ is not.

\subsubsection{The Particle-Mesh Model}
\label{sec_PM}

Under gravitational interaction, the equations of motion
for the mass particles set up above can be written as
\begin{equation}
\frac{{\rm d}{\bf r}}{{\rm d}a}=\frac{1}{a^2\dot{a}}{\bf v}\,;
\end{equation}
\begin{equation}
\frac{{\rm d}{\bf v}}{{\rm d}a}=-\frac{4\pi G\bar\rho_0}{a\dot{a}}\nabla\Phi\,,
\end{equation}
where $\bar\rho_0$ is the mean mass density of the Universe
at $z=0$. Note again that ${\bf v}=a{\bf u}$ with ${\bf u}={a\rm
  d}{\bf r}/{\rm d}t$. The gravitational potential $\Phi$ can be
obtained by solving the Poisson equation,
\begin{equation}
\nabla^2\Phi=\delta(a,{\bf r})\,,
\end{equation}
where $\delta(a,{\bf r})$ is the overdensity field at $z=1/a-1$.

In practice, we use the leapfrog technique to integrate the above
equations forward in time (or in $a$):
\begin{equation}\label{eq_pm4}
{\bf v}({\bf q}, a_{n+1/2})=
{\bf v}({\bf q}, a_{n-1/2})-\nabla\Phi({\bf r}({\bf q}, a_n))
\int^{a_{n+1/2}}_{a_{n-1/2}}\frac{4\pi G\bar\rho_0}{a\dot{a}}{\rm d}a\,;
\end{equation}
\begin{equation}\label{eq_pm5}
{\bf r}({\bf q}, a_{n+1})={\bf r}({\bf q}, a_{n})
+{\bf v}({\bf q}, a_{n+1/2})\int^{a_{n+1}}_{a_{n}}\frac{1}{a^2\dot{a}}{\rm d}a\,.
\end{equation}
For the sake of simplicity, we rewrite these two equations as
\begin{equation}\label{eq_pm6}
{\bf v}_{n+1/2}({\bf q})=
{\bf v}_{n-1/2}({\bf q})+{\bf F}_n({\bf r}_n({\bf q}))\Delta^{\rm v}_n\,;
\end{equation}
\begin{equation}\label{eq_pm7}
{\bf r}_{n+1}({\bf q})=
{\bf r}_{n}({\bf q})+{\bf v}_{n+1/2}({\bf q})\Delta^{\rm r}_n
={\bf r}_{n}({\bf q})+{\bf v}_{n-1/2}({\bf q})
\Delta^{\rm r}_n+{\bf F}_n({\bf r}_n({\bf q}))\Delta^{\rm v}_n\Delta^{\rm r}_n\,,
\end{equation}
where $\Delta^{\rm v}_n$ and $\Delta^{\rm r}_n$ are the integrations
in equations (\ref{eq_pm4}) and (\ref{eq_pm5}), respectively;
$n$ is the step number; and $a_{n+1/2}=(a_n+a_{n+1})/2$.

We adopt the standard procedure to calculate the gravitational
force, ${\bf F}_n({\bf r}_n({\bf q}))$ \citep[e.g.][]{Springel05}.
After obtaining the positions of all mass particles
at the $n$th leapfrog step, ${\bf r}_n({\bf q}_2)$,
we use a clouds-in-cells (CIC) assignment \citep{HockneyEastWood81} to
construct the overdensity field on a grid:
\begin{equation}
\delta_{n, \rm c}({\bf x}_2)=\sum_{{\bf q}_2}w_{\rm c}({\bf x}_2-{\bf r}_n({\bf q}_2))-1\,,
\end{equation}
where $w_{\rm c}({\bf x}_2-{\bf r}_n({\bf q}_2))$ is the CIC kernel in
real space. We Fourier transform the density field and divide it
by the CIC kernel in Fourier space,
$w_{\rm c}({\bf k})={\rm sinc}(k_xL/2N_{\rm c}){\rm sinc}
(k_yL/2N_{\rm c}){\rm sinc}(k_zL/2N_{\rm c})$, to correct for the
smoothing effect of the CIC assignment. The resulting density
field is convolved with a Gaussian kernel $w_{\rm g}( R_{\rm PM}k)$ to
suppress the force anisotropy that may be produced by the finite
size of grid cells. Here $R_{\rm PM}$ is fixed to be $1.2 l_{c}$, with
$l_c=L/N_{\rm c}$ the grid cell size. The smoothed density field at
the $n$th step can then be written as,
\begin{equation}\label{eq_rhon}
\delta_{n}({\bf k})=\frac{w_{\rm g}(R_{\rm PM}k)}
{w_{\rm c}({\bf k})N_{\rm c}^3}
\sum_{{\bf x}_2}{\rm e}^{-i{\bf k}\cdot{\bf x}_2}\delta_{n, \rm c}({\bf x}_2)\,.
\end{equation}
Multiplying $\delta_{n}({\bf k})$ with the Green function,
$-1/k^2$, and with $-i\bf k$, we obtain the gravity in Fourier space.
To obtain accurate forces at particle positions, we first divide the
gravitational force in Fourier space by $w_{\rm c}({\bf k})$, and
then transform it back onto the real space grid:
\begin{equation}
{\bf F}_{{\rm g},n}({\bf x}_1)=
\sum_{\bf k}{\rm e}^{i{\bf k}\cdot{\bf x}_1}\frac{i\bf k}{k^2}
\frac{w_{\rm g}(R_{\rm PM}k)}{w^2_{\rm c}({\bf k})N_c^3}\sum_{{\bf
    x}_2}
\delta_{n, \rm c}({\bf x}_2){\rm e}^{-i{\bf k}\cdot{\bf x}_2}\,.
\end{equation}
We interpolate the forces to particle positions using
a CIC interpolation. Note that the smoothing introduced by the
CIC interpolation has already been de-convolved before the Fourier
transformation. Finally, the gravitational force at a given particle
position can be expressed as
\begin{equation}\label{eq_pmforce}
{\bf F}_n({\bf r}_n({\bf q}_1))=
\sum_{{\bf x}_1}{\bf F}_{{\rm g},n}({\bf x}_1)w_{\rm c}({\bf x}_1-{\bf r}_n({\bf q}_1))\,.
\end{equation}

After $N\equiv N_{\rm PM}$ steps,\footnote{For conciseness, we will use
$N$ to replace $N_{\rm PM}$ in this and the next subsections, and in Appendix \ref{app_hf}.}
we obtain the final particle positions,
${\bf r}_N({\bf q}_1)$, at $z=0$. The final density field is obtained
using the same CIC assignment as described above:
\begin{equation}
\rho_{N, \rm c}({\bf x})=\sum_{{\bf q}_1}w_{\rm c}({\bf x}-{\bf r}_N({\bf q}_1))\,.
\end{equation}
This density field can not yet be used as our final
modeled density field. Although the PM model used here
is much more accurate than many perturbation theories adopted
before, such as the Zel'dovich approximation and
the 2LPT, the accuracy
of the PM result depends on the number of steps adopted,
in the sense that a larger $N_{\rm PM}$ leads to more
accurate results. In practice, however, the value $N_{\rm PM}$ cannot be
chosen to be much larger than 10 in order to complete the HMC
within a reasonable computational time scale. The use of a relatively
small $N_{\rm PM}$ can lead to significant bias in the PM
density field relative to the real density field obtained from a
high-resolution simulation.
Fortunately, this bias can be corrected, at least partly.
Following TZ12, we introduce a density transfer function between the
modeled and real density fields,
\begin{equation}\label{eq_tr}
T(k)=\frac{\langle \rho_{N, \rm c}({\bf k})\rho^{\ast}_{\rm si}({\bf k})
\rangle_{\bf k}}{\langle \rho_{N, \rm c}({\bf k})\rho^{\ast}_{N, \rm c}({\bf k})\rangle_{\bf k}}\,,
\end{equation}
where $\rho_{\rm si}$ is the $z=0$ density field evolved from the
same initial condition as $\rho_{N, \rm c}$ by using an accurate
$N$-body code, here used to represent the real density field.
As demonstrated in TZ12, the density transfer function has small
variance among different realizations (see also Section \ref{sec_tpm}),
and so it is sufficient to estimate it only once. The final model
prediction of the density field we actually use is, in Fourier space,
given by
\begin{equation}\label{eq_rhomod}
\rho_{\rm mod}({\bf k})=\frac{w_{\rm g}(R_{\rm s}k)T(k)}{w_{\rm c}({\bf k})N_{\rm c}^3}\sum_{{\bf x}}
{\rho_{N,\rm c}}({\bf x}){\rm e}^{-i{\bf k}\cdot{\bf x}}\,.
\end{equation}
Note that a new smoothing specified by $R_{\rm s}$,
which is different from $R_{\rm PM}$ used in the PM model,
is introduced here to suppress the shot noise in Hamiltonian
force calculation, and to smooth out the difference between
the modeled and simulated density fields on small scales
produced by the inaccuracy of the dynamic model adopted in the
HMC (see Section \ref{sec_tpm} for details).

\subsection{The Likelihood Term of the Hamiltonian Force}
\label{sec_hf}

As shown in equation (\ref{eq_dp}), the Hamiltonian force consists of two
components, the prior term, $2\delta_j({\bf k})/P_{\rm lin}(k)$, and the
likelihood term, $F_j({\bf k})$. In our model the calculation of the likelihood
($\chi^2$) term consists of three transformations.
The first is the transformation of $\delta_j({\bf k})$ to the initial positions and velocities,
${\bf p}_{\rm i}({\bf q})=[{\bf r}_{\rm i}({\bf q}), {\bf v}_{\rm i}({\bf q})]$,
through the Zel'dovich approximation. The second is the transformation
of the initial positions and velocities set up by the Zel'dovich approximation
to the final positions and velocities
${\bf p}_{N}({\bf q})=[{\bf r}_N({\bf q}),
{\bf v}_{N-1/2}({\bf q})]$
via $N\equiv N_{\rm PM}$ steps of the PM
model.\footnote{It is not necessary to obtain ${\bf v}_N({\bf q})$ because the velocities
are not used in computing $\rho_{\rm mod}$ and $\chi^2$.}
In the following, we use ${\bf p}_{n}({\bf q})=[{\bf r}_n({\bf q}),
{\bf v}_{n-1/2}({\bf q})]$ ($n=1,2,...,N$) to indicate the positions
and velocities of particles at the $n$th PM step. Sometimes we also
use ${\bf p}_{0}({\bf q})=[{\bf r}_0({\bf q}), {\bf v}_0({\bf q})]$
to indicate the initial condition, i.e. ${\bf p}_{\rm i}({\bf q})$.
The third one is the transformation of the final positions to
the $\chi^2$. The chain rule of differentiation allows us to rewrite
the derivatives of $\chi^2$ with respect to $\delta_j({\bf k})$ as
\begin{equation}\label{HforcesCal}
F_j({\bf k}) = \frac{\partial\chi^2}{\partial
\delta_j({\bf k})}=
\frac{\partial\chi^2}{\partial {\bf p}_N}
\otimes\frac{\partial{\bf p}_N}{\partial {\bf p}_{N-1}}
\otimes\cdot\cdot\cdot
\otimes\frac{\partial{\bf p}_2}{\partial{\bf p}_{1}}
\otimes\frac{\partial{\bf p}_1}{\partial {\bf p}_{\rm 0}}
\otimes\frac{\partial{\bf p}_{\rm 0}}{\partial\delta_j({\bf k})}\;.
\end{equation}
Here $\partial{\bf p}_{n+1}/\partial {\bf p}_{n}$
($n={\rm i},1,2,\cdot\cdot\cdot, N-1$) is a $6\times 6$ (three
coordinates and three components of velocity),
$\partial\chi^2/\partial {\bf p}_N$ a $1\times 6$, and
$\partial{\bf p}_{\rm 0}/\partial\delta_j({\bf k})$ a $6\times 1$
matrix, and $\otimes$ denotes matrix multiplication.

As suggested by \citet{HansonCunningham96}, the matrix
multiplications in the above equation can be carried out
in two different ways. The first proceeds in the same
time sequence as the structure formation, corresponding
to the inverse order of the right hand side of
equation (\ref{HforcesCal}). Namely one first calculates
$\partial{\bf p}_{\rm 0}/\partial\delta_j({\bf k})$, then
$\partial{\bf p}_{1}/\partial\delta_j({\bf k})$, and so on
to $\partial{\bf p}_{N_{\rm PM}}/\partial\delta_j({\bf k})$,
and eventually obtaining $\partial\chi^2/\partial {\delta_j({\bf k})}$.
However, calculation along this sequence results in
very large intermediate matrices -- for example
$\partial{\bf p}_{\rm 0}/\partial\delta_j({\bf k})$ has
$N_{\rm c}^6$ variables, and is almost impossible to
handle.   The second way proceeds in the opposite direction.
One first calculates $\partial\chi^2/\partial {\bf p}_{N}$,
then $\partial\chi^2/\partial {\bf p}_{N-1}$, and so on
to $\partial\chi^2/\partial {\bf p}_{0}$ (which by definition is
equal to  $\partial\chi^2/\partial {\bf p}_{\rm i}$),
eventually obtaining $\partial\chi^2/\partial {\delta_j({\bf k})}$.
This technique is called the adjoint differentiation technique
by \citet{HansonCunningham96}.

In this order the calculation of the likelihood term of the
Hamiltonian force consists of the following three parts.
The first is the $\chi^2$ transformation, which calculates
$\partial\chi^2/\partial {\bf r}_{N}({\bf q})$ using the relation
between $\rho_{\rm mod}$ and ${\bf r}_N$. Note that
$\partial\chi^2/\partial {\bf v}_{N-1/2}({\bf q})\equiv 0$
since particle velocities are not used in our definition of
$\chi^2$. The second is the Particle-Mesh transformation,
which obtains $\partial\chi^2/\partial {\bf r}_{n}({\bf q})$
and $\partial\chi^2/\partial {\bf v}_{n-1/2}({\bf q})$ through
the values of $\partial\chi^2/\partial {\bf r}_{n+1}({\bf q})$
and $\partial\chi^2/\partial {\bf v}_{n+1/2}({\bf q})$ obtained
in a previous step, using the relations between
$[{\bf r}_n ({\bf q}),  {\bf v}_{n-1/2}({\bf q})]$ and
$[{\bf r}_{n+1} ({\bf q}),  {\bf v}_{n+1/2}({\bf q})]$ given
by the PM model. Finally, the Zel'dovich transformation
relates $[{\bf r}_{\rm i}({\bf q}),  {\bf v}_{\rm i}({\bf q})]
=[{\bf r}_0 ({\bf q}),  {\bf v}_0({\bf q})]$ to
$\delta_j ({\bf k})$.  We introduce a transitional matrix,
\begin{equation}
{\bf \Psi}({\bf q})=N_{\rm c}^3
\left[\frac{\partial\chi^2}{\partial{\bf r}_{\rm i}({\bf q})}
+\frac{\partial\chi^2}{\partial{\bf v}_{\rm i}({\bf q})}
H_{\rm i}a^2_{\rm i}f(\Omega_{\rm i})\right]\,,
\end{equation}
and  derive the expression of the likelihood term of the
Hamiltonian force for the real part of $\delta({\bf k})$ as
\begin{equation}\label{eq_f0}
F_{\rm re}({\bf k})=\frac{2D(a_{\rm i})}{k^2}{{\bf k}\cdot{\bf\Psi}_{\rm im}({\bf k})}\,,
\end{equation}
and for the imaginary part as
\begin{equation}\label{eq_f1}
F_{\rm im}({\bf k})=-\frac{2D(a_{\rm i})}{k^2}{{\bf k}\cdot{\bf\Psi}_{\rm re}({\bf k})}\,.
\end{equation}
Here ${\bf\Psi}_{\rm re}({\bf k})$ and ${\bf\Psi}_{\rm im}({\bf k})$
are, respectively, the real and imaginary parts of ${\bf\Psi(k)}$, the Fourier
transform of ${\bf \Psi}({\bf q})$.

The details of the calculation of all the terms in the likelihood
term of the Hamiltonian force are given in Appendix \ref{app_hf}.

\section{$N$-body Simulations and the Performance of the PM model}
\label{sec_sim}

\subsection{$N$-body Simulations}

In this paper, we use four $N$-body simulations to test the performance
of our HMC+PM method. These simulations are obtained using
Gadget-2 \citep{Springel05}. The initial conditions are set up using
the method presented in Section \ref{sec_za}. Two of them, which
are referred to as L300A and L300B, assume a spatially flat
$\Lambda$CDM model, with the present density parameter
$\Omega_{\rm m,0} = 0.258$, the cosmological constant
$\Omega_{\Lambda,0}= 0.742$ and the baryon density parameter $\Omega_{\rm b,0}= 0.044$,
and with the power spectrum obtained using the linear transfer function
of \citet{EisensteinHu98} with an amplitude specified by $\sigma_8= 0.80$.
The CDM density field of each simulation was traced
by $512^3$ particles in a cubic box with a side length of 300 $\mpc$. The other two simulations, referred
to as L100A and L100B, assume the same cosmological model
as the L300 simulations, but use $512^3$ particles to trace the
evolution of the cosmic density field in a smaller, 100 $\mpc$ box.
The initial redshifts for the L300 and L100 simulations are
set to be 36 and 72, respectively. For the cosmological model adopted here, the characteristic nonlinear scale at
the present is about $7\mpc$, corresponding to a wave-number
$k\simeq 0.15\mpci$.

\subsection{Parameters and Performance of the PM Model}\label{sec_tpm}

The accuracy and reliability of the HMC method relies on the adopted model of structure evolution.
This is the main reason why we propose to employ the PM model, instead of other
simpler models with lower accuracy, to evolve the cosmic density field.
In principle, the PM model can yield a density field with high precision,
as long as the grid cell size, $l_{\rm c}=L/N_{\rm c}$, and the time step,
characterized by $\log(1+z_{\rm i})/N_{\rm PM}$, are chosen to be sufficiently
small. In practice, however, it is feasible only to adopt a finite
cell size and a finite time step, because of limited computational resource.
It is thus necessary to explore the dependence of the quality of the PM model
on $l_{\rm c}$ and $N_{\rm PM}$, which can then help us to properly set
the parameters in our HMC. One way to quantify the quality
of a structure evolution model is to measure its similarity to
the density field obtained from a high-resolution simulation
with the same initial condition. The similarity between any two density fields,
$X$ and $Y$, can be quantified by using the phase correlation of their
Fourier transforms \citep[see e.g. TZ12;][]{KitauraHb13},
\begin{equation}
C_{p}(k, X, Y) = \frac{\langle X({\bf k})Y^{\ast}({\bf
k})\rangle _{\bf k}}{\sqrt{\langle
|X({\bf k})|^2\rangle_{\bf k}\langle |Y({\bf k})|^2\rangle_{\bf
k}}}\,.
\label{eq_ph}
\end{equation}
Evidently, $C_{p}(k)=1$ implies that the two fields have
exactly the same phase, while $C_{p}(k)=0$ indicates null correlation,
for the Fourier mode in question.

\begin{figure*}
\centering
\epsfig{file=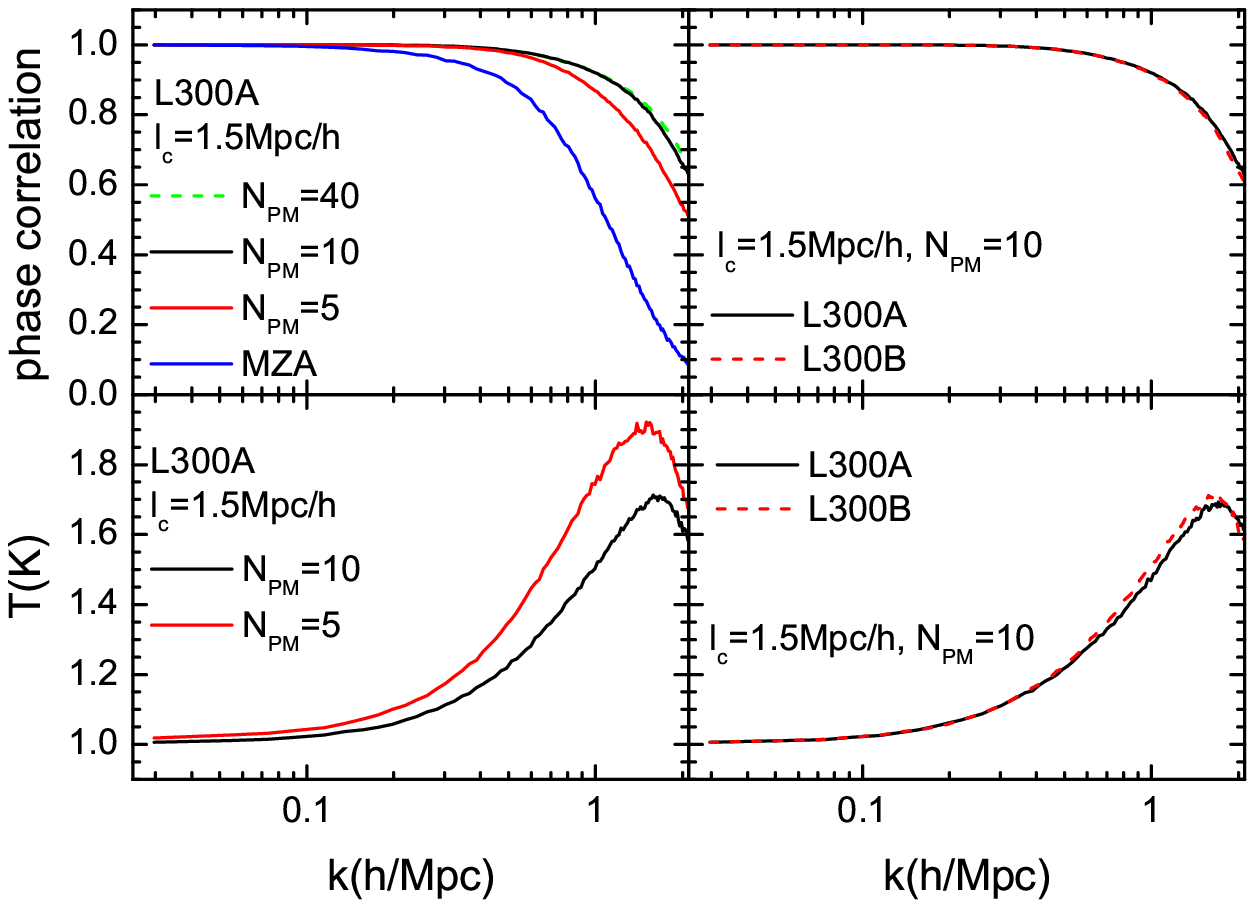,scale=1.}
\caption{Upper left panel: the phase correlations between $N$-body simulated density field (L300A)
and various modeled density fields as indicated in the panel. Here $N_{\rm PM}$ is the number
of steps adopted in a PM model and $l_{\rm c}$ is the grid cell size. Upper right panel: the
phase correlations between the simulated density fields and the corresponding PM density fields
with $N_{\rm PM}=10$. Lower left panel: the density transfer functions for PM models
with $N_{\rm PM}=5$ and 10. Lower right panel: density transfer functions for PM model
with $N_{\rm PM}=10$, derived from two simulations. The grid cell size is set to $1.5\mpc$ for
all cases.}
\label{fig_phasetr}
\end{figure*}

As an example, we show in Figure \ref{fig_phasetr} such
phase correlation for a PM model using $l_{\rm c}=1.5\mpc$.
The upper right panel shows the phase correlations
between the two L300 simulations and the corresponding PM density fields
at redshift zero. The number of PM steps ($N_{\rm PM}$) used here
is 10.  The two curves are almost identical, suggesting that the
quality of the PM model is insensitive to sample variance.
At large scales, the PM density fields almost perfectly match the
original simulations. Even at the highly non-linear scale
$k\sim 2.0\mpci$, the correlation coefficient is still larger
than $0.6$. For comparison, the correlations for PM models
using $N_{\rm PM}=5$, $10$ and $40$ are shown in the
upper left panel. Here results are presented only for L300A.
As expected, the quality of the PM model improves with
increasing $N_{\rm PM}$. The improvement is large at
$N_{\rm PM}\leq10$, but becomes saturated at $N_{\rm PM}>10$.
The reason is that once the typical motion of particles in
one time step is much less than the grid cell size, as is the case
for $N_{\rm PM}=40$ and $l_{\rm c}=1.5\mpc$, the use of
a higher time resolution becomes unnecessary.

\begin{figure*}
\centering
\epsfig{file=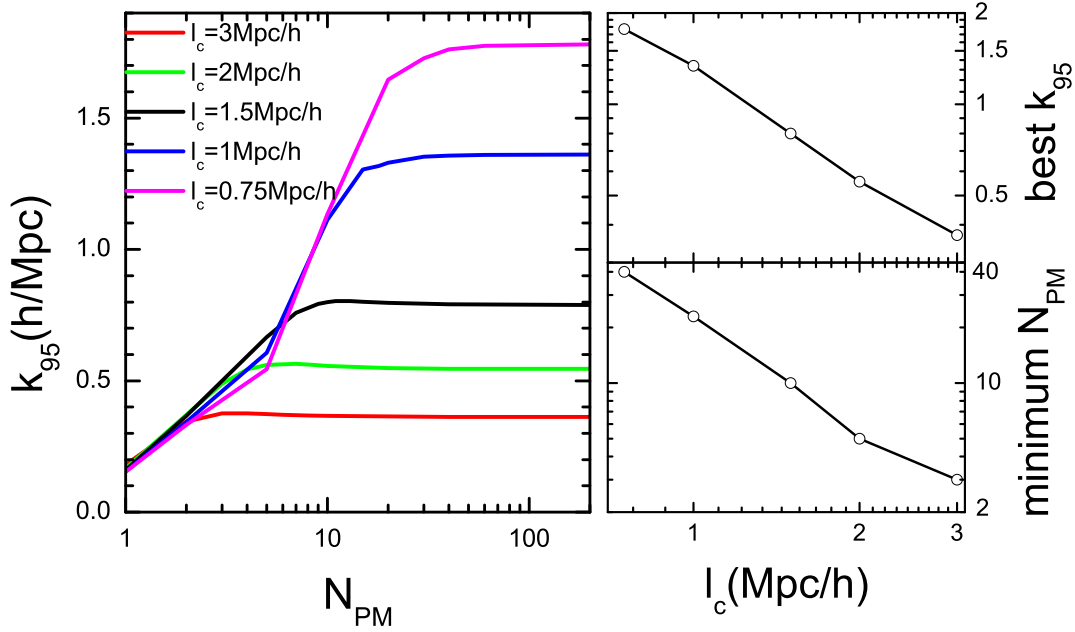,scale=1.}
\caption{Left panel: the parameter $k_{95}$ as a function of $N_{\rm PM}$ for different
grid cell sizes, $l_{\rm c}$, as indicated in the figure. Here $k_{95}$ measures the scale
at which the phase correlation between the PM density field and the original
simulated density field is 0.95. Right panels: the best $k_{95}$ and the required
minimum $N_{\rm PM}$ as a function of $l_{\rm c}$. All of the results
are based on the simulation L300A.}
\label{fig_Npmlc}
\end{figure*}

In order to further characterize the phase correlation, we introduce
a quantity, $k_{95}$, which is defined to be the wavenumber
at which the correlation coefficient defined in equation\,(\ref{eq_ph})
decreases to $0.95$. A larger $k_{95}$ therefore indicates a more
accurate model prediction. We show $k_{95}$ as a function of
$N_{\rm PM}$ for various $l_{\rm c}$ in the left panel of
Figure \ref{fig_Npmlc}. For all the cell sizes examined,
$k_{95}$ first increases with $N_{\rm PM}$, and then remains
at an almost constant level. For a given grid cell size $l_c$,
there is thus an upper limit in the performance of the PM model,
consistent with the results shown in Figure \ref{fig_phasetr}.
Of course, the level of the best performance increases
as $l_c$ decreases. For instance, the value of $k_{95}$
for $l_{\rm c}=3$, $1.5$ and $0.75\mpc$ are about $0.38$, $0.80$
and $1.78\mpci$, respectively (see the upper right panel of
Figure \ref{fig_Npmlc}), roughly proportional to $1/l_{\rm c}$.
Moreover, these results suggest that, for a given $l_c$, there is
a minimum $N_{\rm PM}$ such that the performance of the PM model
almost reaches its best and any further increase in
$N_{\rm PM}$ no long makes significant improvement.
This minimum $N_{\rm PM}$ roughly scales
as $1/l_{\rm c}^{2}$, as shown in the lower right panel of
Figure \ref{fig_Npmlc}, and is about $3$, $10$ and
$40$ for $l_{\rm c}=3$, $1.5$ and $0.75\mpc$, respectively.

For given $N_{\rm PM}$ and $l_{\rm c}$, significant bias can exist
in the PM density field relative to the real density field.
In order to correct for this effect, following TZ12, we
introduce a density transfer function, $T(k)$,  which is
obtained by cross-correlating the original simulation and
the PM density field [equation (\ref{eq_tr})]. Note that introducing
$T(k)$ into the calculation of the final density field does not change
the phase correlation shown above. As an example, the lower left
panel of Figure \ref{fig_phasetr} shows the transfer function for
$l_{\rm c}=1.5\mpc$ and $N_{\rm PM}=5$ or $10$.
As expected, $T(k)$ is almost unity at large scales and
increases gradually with increasing $k$ until $k\sim1.5\mpci$.
An downturn is observed in the transfer functions at small
scales, indicating a rapid decline in the correlation between
the two density fields at such scales. Since the PM model with
$N_{\rm PM}=10$ (referred to as PM10) is more accurate than
PM5, the transfer function for the PM10 model is
closer to unity than that for the PM5 model.  Similar to the phase
correlations, the transfer functions are insensitive to sample variance
(see the lower right panel of Figure \ref{fig_phasetr}). In what follows,
PM model really means PM model combined with the corresponding
transfer function.

\begin{figure*}
\centering
\epsfig{file=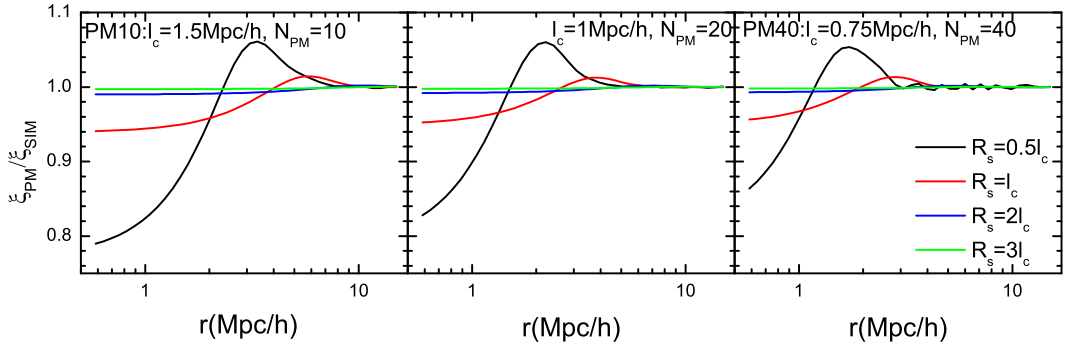,scale=1.}
\caption{The ratios between the two-point correlation functions measured from the PM
and the simulated density fields for three PM models as indicated in the panels.
The density fields are smoothed with various smoothing scales, $R_{\rm s}$, as indicated.}
\label{fig_2pc}
\end{figure*}

The other way to reduce the bias produced by
the inaccuracy of the PM model at small scales is to
smooth both the PM density field and the input density field
at small scales before using them to calculate the likelihood
function in the HMC.
Here we investigate which smoothing scale,
$R_{\rm s}$, is suitable for our purpose.
To this end, we calculate the two point correlation functions for
both the PM density field ($\xi_{\rm PM}$) and the original,
simulated density field ($\xi_{\rm SIM}$), both smoothed with the
same smoothing scale. Figure \ref{fig_2pc} shows the ratio,
$\xi_{\rm PM}/\xi_{\rm SIM}$, for a number of smoothing scales.
The results are shown for three PM models, one with $l_{\rm c}=1.5\mpc$
and $N_{\rm PM}=10$, another with
$l_{\rm c}=1\mpc$ and $N_{\rm PM}=20$, and the third with
$l_{\rm c}=0.75\mpc$ and $N_{\rm PM}=40$ (hereafter referred to as PM40). We note again
that all the density fields are corrected with the density transfer
functions. The model predictions are in good agreement with the
original ones at large scales for all models, independent of
the smoothing scale adopted. At small scales, however,
the PM density fields are less clustered than the simulated ones.
This discrepancy is significant only when the smoothing scale,
in units of $l_{\rm c}$, is chosen to be too small,
but almost vanishes for $R_{\rm s}\ge 2l_{\rm c}$. This basically says
that PM density field is inaccurate on scales below a couple of grid
cells and the difference between the model prediction
and the input field on such scales should be suppressed
with smoothing. However, choosing a too large $R_{\rm s}$ will
cause loss of information on small scales. As a compromise we will
use $R_{\rm s}= 3l_{\rm c}$ as our fiducial value.

\section{Applications and Tests of the HMC+PM Method}
\label{sec_app}

\subsection{Setting Parameters for the HMC}
\label{sec_phmc}

Before describing the applications of our PM based HMC (HMC+PM), we briefly describe how to specify the parameters in the HMC and
in the PM model.
We divide the simulation boxes into $N^3_{\rm c}$ grid cells and use
a Gaussian kernel with a smoothing scale of $R_{\rm s}$ to smooth the
particle distributions at redshift zero on to the grids. The resultant density
fields, denoted by $\rho_{\rm si}(z=0)$, are what we want to match in the
reconstructions [i.e. $\rho_{\rm inp}=\rho_{\rm si}(z=0)$].

Our dynamic model consists of two parts.
First, the Zel'dovich approximation has one parameter, the initial redshift $z_{\rm i}$.
In the present paper we set $z_{\rm i}=36$, the same as that for the initial
conditions of the L300 simulations. Second, the PM model (including
the transfer function) has two parameters, the grid cell size,
$l_{\rm c}$, and the number of time steps, $N_{\rm PM}$.
The computation time for a PM model is proportional to
$N_{\rm PM}$ and $1/l^3_{\rm c}$. As a compromise between
computation time and model precision, we adopt two PM models,
PM10 with $l_{\rm c}=1.5\mpc$ and $N_{\rm PM}=10$, which is
implemented in the reconstructions of the L300 series;
PM40 with $l_{\rm c}=0.75\mpc$ and $N_{\rm PM}=40$, used in
the reconstructions of the L100 series. The corresponding
$N_{\rm c}$ is $200$ for L300 and $134$ for L100.

According to the tests shown above, we choose the smoothing
scale $R_{\rm s}=3l_{\rm c}$ in our HMC runs to make sure that
the bias in the PM density field on small scales is sufficiently
suppressed. A series of tests were done
in W13 to tune other HMC parameters. Following their results,
we adopt $n_{\rm max}=13$, $\tau_{\rm max}\approx0.1$
and set $\sigma_{\rm inp}({\bf x})=\mu\rho_{\rm si}({\bf x})$
with $\mu=0.5$. The weight field $w({\bf x})$ is set to be
unity for all grid cells.

\subsection{Applications to Simulated Density Fields}
\label{sec_asd}

\begin{figure*}
\centering
\epsfig{file=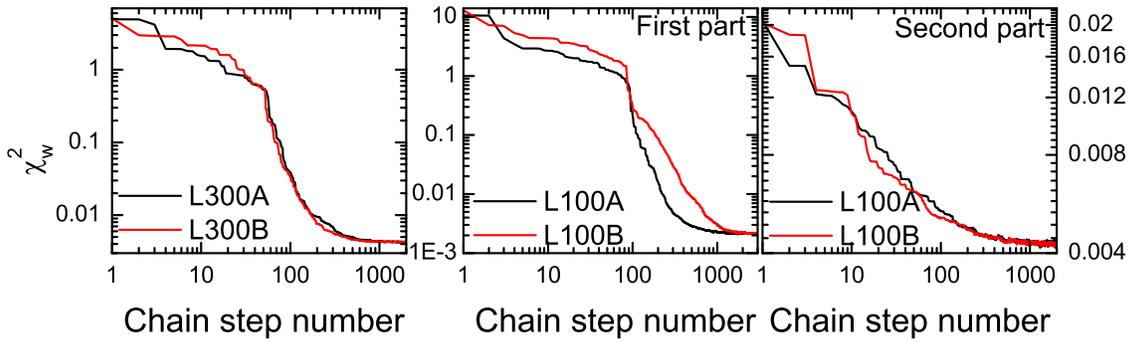,scale=1.}
\caption{$\chi^2_{w}=\chi^2/\sum_{\bf x}w({\bf x})$ as a function of chain step for four HMC
reconstructions. The left panel shows the results for the L300 series, while the
middle and right panels show the results for the first and second HMC parts for
the L100 series. The first HMC part adopts a larger smoothing scale than the second part.}
\label{fig_chi2}
\end{figure*}

We use the four $N$-body simulations, L300A, L300B, L100A
and L100B, to test our HMC+PM method. Note again that PM10
and PM40 are adopted for the reconstructions of the
L300 and L100 series, respectively. The input density field
$\rho_{\rm inp}=\rho_{\rm si}(z=0)$ is smoothed with a smoothing scale
$R_{\rm s}=n_{\rm s}l_{\rm c}$, and $n_{\rm s}$ are set to
be 3 and 4 for L300 and L100, respectively.
The initial set of $\delta({\bf k})$ is randomly drawn from the
prior Gaussian distribution, $G(\delta({\bf k}))$, as shown in
equation (\ref{eq_post}). The Hamiltonian masses are computed
twice during the entire process. The first time is at the
beginning. After proceeding 50 or 80 accepted chain steps,
the mass variables are updated once with the current
Hamiltonian forces and retained all the way to the end.
Computations are made for 2000 (3000) HMC steps for the
L300 (L100) series, and the acceptance rates for the reconstructions
of the L300 and L100 series are about 83\% and 96\%, respectively.
The left and middle panels of Figure \ref{fig_chi2} show
$\chi^2_{w}\equiv \chi^2/\sum_{\bf x}w({\bf x})$ as a function
of the chain step. The value of $\chi^2_w$ drops sharply at the
beginning (the burn-in phase), and then remains almost constant
after about $500$ - $1000$ chain steps (the convergence phase).
The $\chi^2_w$ values of the converged steps are about 0.004
(L300) and 0.002 (L100), showing that the predicted density
fields after convergence match the input ones well.

For the reconstruction of the L100 series, further HMC process is
carried out to increase the reconstruction accuracy. This time
the input density field is still $\rho_{\rm si}$ but smoothed with
a smaller smoothing scale, $R_{\rm s}=3l_{\rm c}=2.25\mpc$.
The initial set of $\delta({\bf k})$ is not randomly generated,
but taken from the linear density field output at the 2700th step
of the first HMC process, and so are the Hamiltonian masses.
The new fictitious systems are then evolved for additional 2000 steps,
and the acceptance rates here are about 95\%. The right panel of
Figure \ref{fig_chi2} shows $\chi^2_{w}$ as a function of chain
step for the additional runs. At the first step, $\chi^2_{w}$ is about 0.02, much larger than the final one of the first HMC run. It is ascribed to the smaller smoothing scale adopted here.
One can see that two-phase behavior is conspicuous too.
After a quick decline within the first 200 steps, $\chi^2_w$
converges to about 0.004. Since the initial set of $\delta({\bf k})$ is
not random, the decreasing in the amplitude of $\chi^2_{w}$ in
the second part of the HMC is much slower. We note that
the two-phase behavior is commonly seen in HMC runs \citep[e.g.][]{Hanson01,Taylor_etal08}.
As discussed in W13, the fictitious system in question
actually mimics a cooling system in a gravitational potential well.
The reset of momenta at the beginning of every chain step
(see Section \ref{sec_HMCMC}) is an analog to the `cooling' process,
which makes the system fall continuously and eventually reach
the bottom region of the potential well (i.e. around the posterior
peak we are searching for).

\begin{figure*}
\centering
\epsfig{file=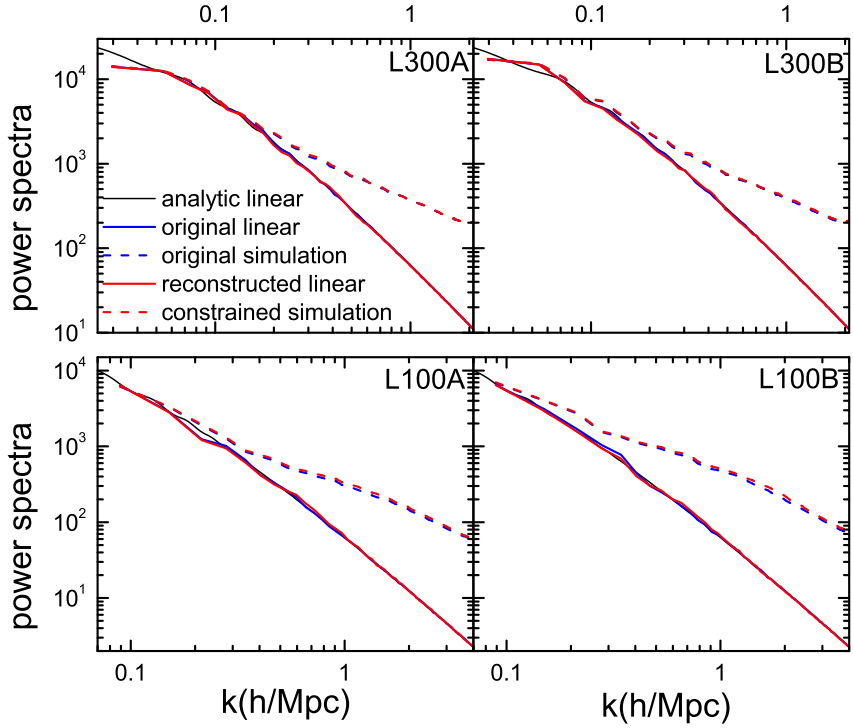,scale=1.}
\caption{The blue and red solid lines show the original and reconstructed
linear power spectra. The blue and red dashed lines show the $z=0$ power
spectra measured from the original input simulations and corresponding CSs.
The black solid lines show the analytic linear power spectrum.}
\label{fig_ps}
\end{figure*}

On the scale of 4.5$\mpc$ (2.25$\mpc$), the {\it RMS} (root mean square)
difference between $\rho_{\rm inp}$ and $\rho_{\rm mod}$ is about
$\mu\sqrt{2\chi^2_w}\simeq4.5\%$ ($4.6\%$) for the L300 (L100) series.
This is an accurate match, indicating that the second constraint
in our reconstruction, namely that $\rho_{\rm mod}$ matches
$\rho_{\rm inp}$, is well satisfied. In order to check whether our
reconstruction also meets the first constraint, i.e., the prior Gaussian
distribution with a given linear power spectrum, we show in
Figure \ref{fig_ps} the power spectra measured from the
reconstructed linear density fields at the final chain steps,
with the original linear spectra overplotted for comparison.
Over the entire range of wavenumbers, the reconstructed
initial spectra are in good agreement with the original ones.
Examining closely,  however, we see some small deviations.
For example, the reconstructed linear spectra contain
a small bump (less than $10\%$) at the intermediate scale
($k\sim 0.9\mpci$ in the L100 series). Similar but significantly
stronger bump is also seen in the reconstructed linear
spectra of W13, \citet[their figure 2]{Kitaura13} and \citet{Ata_etal14}.

In order to understand the origin of this discrepancy, we
apply our HMC$+$PM reconstruction with PM density fields
as input. Since the input field is generated with the same
PM model as that used in the HMC, no bias is expected from
the inaccuracy of the PM model. As shown in
Appendix \ref{app_apd}, even in such cases, there
is still deviation on scales $\sim R_s$, below which
the prior term of the Hamiltonian force starts to dominate
over the likelihood term. Fortunately, the deviation is
quite small (typically less than $10\%$) and can be
moved to small scales of no practical importance by adopting a
sufficiently small $l_c$ (so that $R_{\rm s}$ is sufficiently
small).

\begin{figure*}
\centering
\epsfig{file=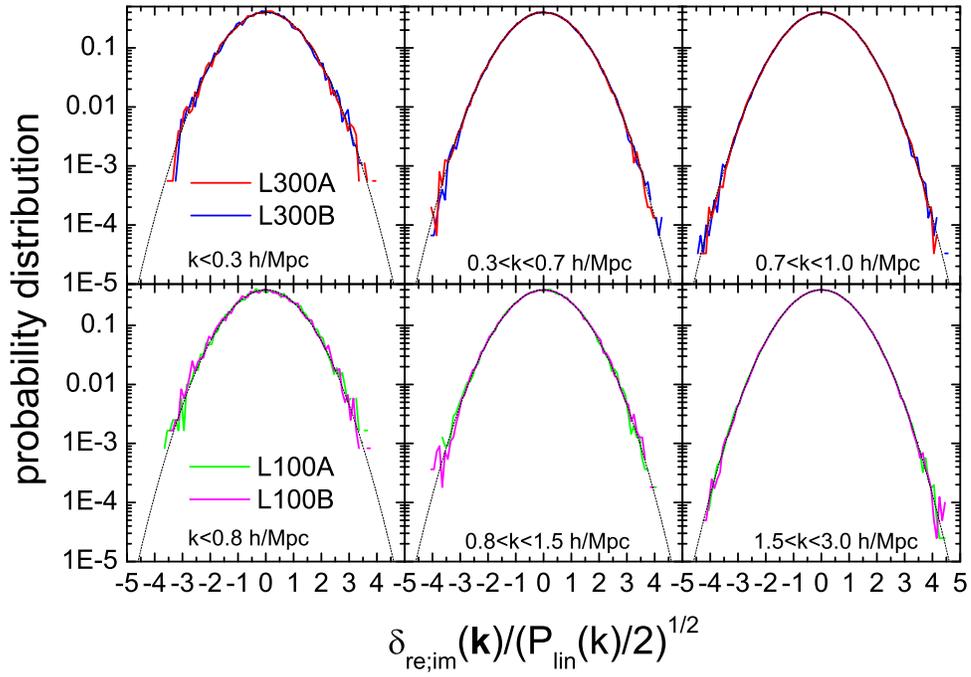,scale=1.}
\caption{The distributions of $\delta_{j}({\bf k})/\sqrt{P_{\rm lin}(k)/2}$ at three different scales as indicated in the panels. Here $\delta({\bf k})$ is the reconstructed linear density field. The smooth curves
are Gaussian distributions with dispersion $\sigma=1$. }
\label{fig_gaudis}
\end{figure*}

For a random Gaussian field,
$d_{{\rm n},j}({\bf k})\equiv \delta_j({\bf k})/\sqrt{P_{\rm lin}(k)/2}$
should obey a Gaussian distribution with $\sigma=1$,
independent of the wavenumber $k$. Figure \ref{fig_gaudis} shows
the distributions of $d_{{\rm n},j}({\bf k})$  for three different
wave-numbers, together with a Gaussian function with
$\sigma=1$. These distributions match the Gaussian distribution
very well even in the large $|d_{{\rm n},j}({\bf k})|$ tails. The fluctuations
at the tails are due to small number statistics.  We have also
calculated the third- and fourth-order moments of the distributions
and estimated the corresponding skewness and kurtosis
\citep[using equations 6.43 and 6.44 in][]{Mobook10}.
Both are found to be very close to zero. All these demonstrate
that our reconstruction recovers the Gaussianity of the linear
density field very well.

At small scales, the Hamiltonian forces are dominated by the
prior term, as the likelihood term is strongly suppressed by the
smoothing (see Appendix \ref{app_apd}). Consequently the phases
of the reconstructed $\delta({\bf k})$ are expected to be
random on small scales. To quantify on which scales
the reconstructed $\delta({\bf k})$ matches the original ones well, we measure the
phase correlation between the two linear density fields. The
results are shown as solid lines in the two left panels of
Figure \ref{fig_phasers}. For the L300 series, the correlation
coefficients are larger than $0.95$ at $k<0.28\mpci$, and
decline gradually to $0.5$ at $k\sim 0.47\mpci$. For the L100
series the phase correlation is even better, with the correlation
coefficient reaching $0.95$ at $k\sim 0.36\mpci$ and
declining to $0.5$ at $k\sim 0.85$. The correlation is still
significant ($C_p\sim 0.1$) at $k\sim1.3\mpci$. The improvement
is clearly due to the better PM model that can
recover better the phase information on small scales. Note that
the results of the two reconstructions in the same series are
very similar, demonstrating again the robustness of our HMC+PM
method.

So far we have shown the results of the linear density field
obtained from the final chain step of each reconstruction. We have
also checked the results at other steps after burn-in and found that they
have almost the same statistical properties as the final chain step.
Because our goal here is to reconstruct $\delta({\bf k})$ rather than
to draw a posterior ensemble of linear density fields, we will not
present the results for chain steps other than the final one.

\section{N-body simulations of the reconstructed initial density fields}
\label{sec_cs}

To further investigate the accuracy of our method, we use the reconstructed
linear density fields to set up initial conditions, and evolve them to
the present day with the $N$-body code Gadget-2 \citep{Springel05}.
These new simulations are referred to as constrained simulations (CS).
The initial conditions are sampled with the same number of
$N$-body particles as in the corresponding original simulations.
The Nyquist frequency for these initial conditions is larger than
$k_{\rm rc}=N_{\rm c}\pi/L$, the largest working frequency of
our HMC method. We complement the Fourier modes at
$k>k_{\rm rc}$ by sampling $\delta({\bf k})$ from the prior Gaussian
distribution, $G(\delta({\bf k}))$.  In the following a suffix `-CS' is
added after the name of an original simulation to denote the
corresponding CS.  For example, L100A-CS denotes the CS of
L100A. The corresponding density field of a CS is denoted
by $\rho_{\rm cs}$.

\begin{figure*}
\centering
\epsfig{file=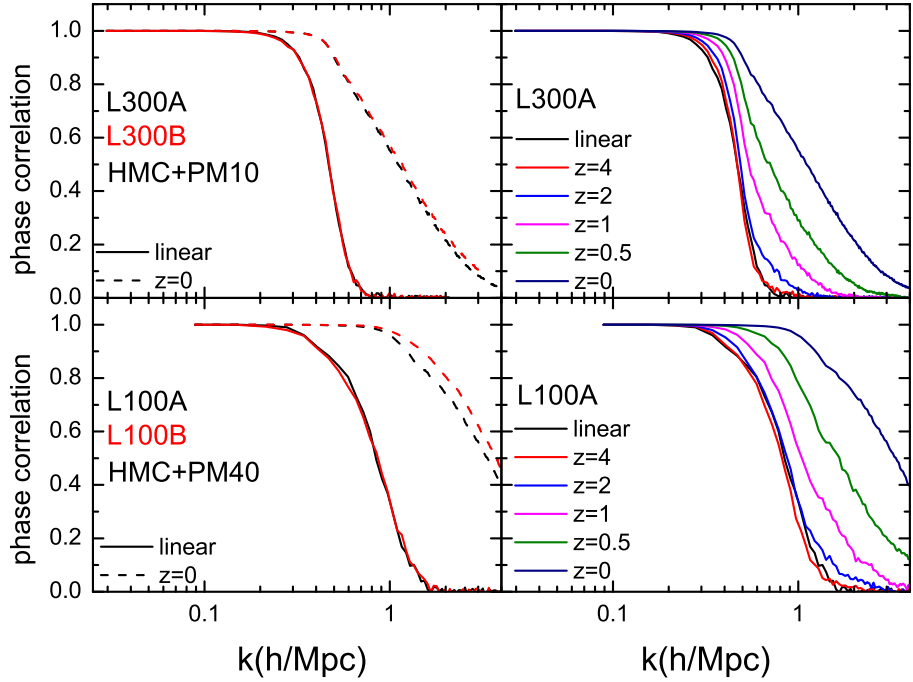,scale=1.}
\caption{Left panels: the solid lines are the phase correlations between the original
and reconstructed linear density fields, while the dashed lines are the phase correlations
between the original (input) simulations and the corresponding CSs at
redshift zero. Right panels: the phase correlations between the original simulation
and the CS at various redshifts. The black line is the correlations of the linear density fields.}
\label{fig_phasers}
\end{figure*}

The CS power spectra at redshift zero are presented in Figure \ref{fig_ps}.
As can be expected from the good accuracy in the reconstructed linear
density fields,  the CS power spectra match their original counterparts
well. The phase correlations between $\rho_{\rm cs}$ and $\rho_{\rm si}$
at $z=0$ are shown as dashed lines in the left two panels of
Figure \ref{fig_phasers}. The correlations are very tight. For example,
the L100-CSs almost perfectly match the original ones all the way
to non-linear scales, $k\sim1.0\mpci$. Even at highly
non-linear scales ($k\sim3.4\mpci$), about half of the phase
information is recovered. The correlations are much stronger than
that obtained from the HMC$+$MZA method. The latter predicts
a more rapid decrease of the correlation function with
increasing $k$, reaching $0.5$ at $k\sim1\mpci$
(W13). The improvement
of our HMC$+$PM over the HMC$+$MZA is clearly due the
more accurate PM model used in the HMC$+$PM for evolving
the cosmic density field.
The phase correlation between  $\rho_{\rm cs}$
and $\rho_{\rm si}$ is also much tighter than that of the
corresponding linear density fields (solid lines). For instance, at $k=1\mpci$
the correlation between linear density fields is only modest for the L100
series, in contrast to the almost perfect
correlation between the fully evolved density fields
(dashed lines). This phenomenon can be readily interpreted by
non-linear mode coupling, in which  the small-scale power of a
non-linear density field is partly produced by the large scale
power \citep[e.g.][]{TassevZaldarriaga12a}. Thus, even
phase correlation is absent on small scales between the
reconstructed and original linear fields, it can be generated
by the phase correlation on large scales that is present in the
linear fields.

Since non-linear effect is more important at lower redshift,  the
mode coupling effect is expected to be less significant at high $z$.
To demonstrate this, we show the phase correlations between
$\rho_{\rm cs}$ and $\rho_{\rm si}$ at five different redshifts, together
with that for the initial density fields, in the right panels of
Figure \ref{fig_phasers}. Take the results for L300A as an example
(the upper right panel). The correlation at $z=4$ is very similar
to that of the linear field, indicating that mode coupling has not
taken its effect.  At $z=2$, the correlation coefficient is
enhanced significantly at $k\sim0.9\mpci$, but remains
unchanged at $k\sim0.5\mpci$. As the evolution proceeds
to $z=0.5$, the effect becomes visible over the entire range of
wave-numbers, where the initial phases are not well constrained.
Similar behavior is also found for L100A, as shown in the lower
right panel. This demonstrates again that an accurate dynamic model of structure evolution is required in order to have accurate
reconstruction in the non-linear regime.

\begin{figure*}
\centering
\epsfig{file=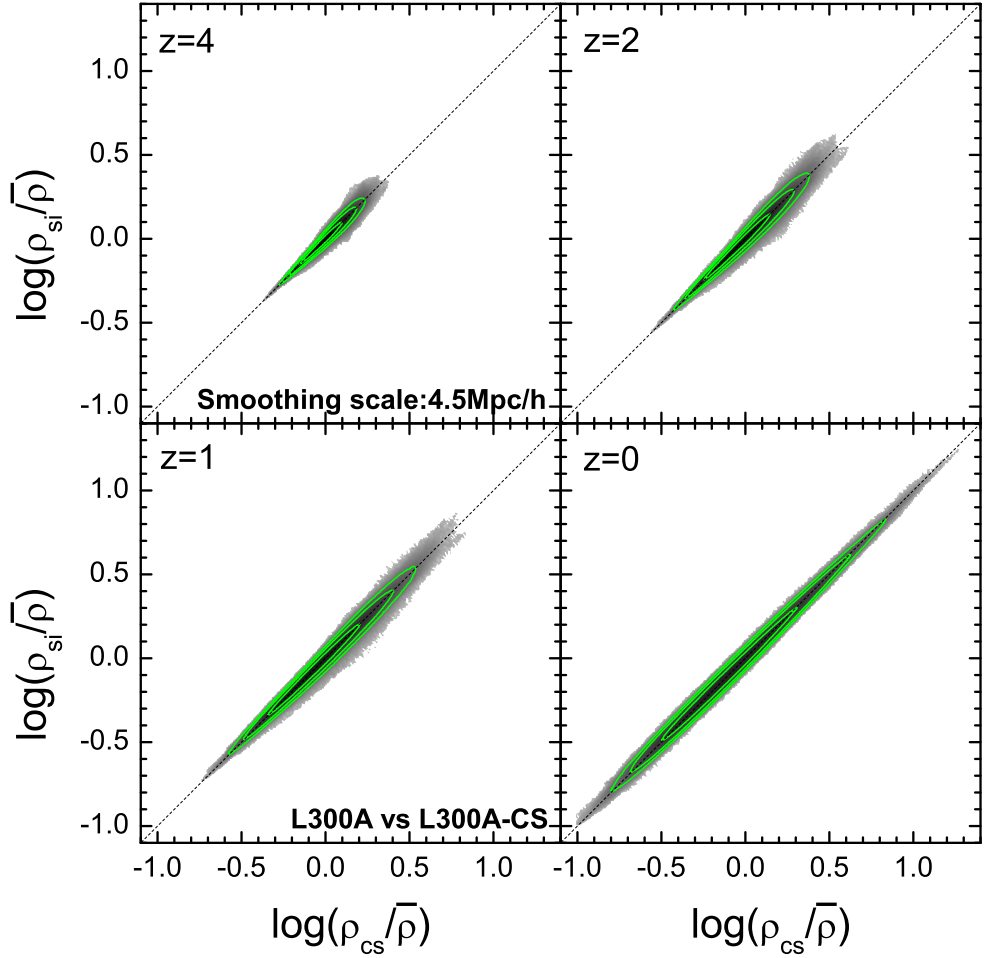,scale=1.}
\caption{The density-density plots between the original simulation, L300A, and
the corresponding CS at various redshifts. The density fields are smoothed with
a Gaussian of radius $3\mpc$. The three contours encompass
67\%, 95\% and 99\% of all the grid cells in the simulation box. All these
densities are scaled with $\bar\rho$, the mean density of the universe at the
redshift in question.}
\label{fig_L300rho}
\end{figure*}

\begin{figure*}
\centering
\epsfig{file=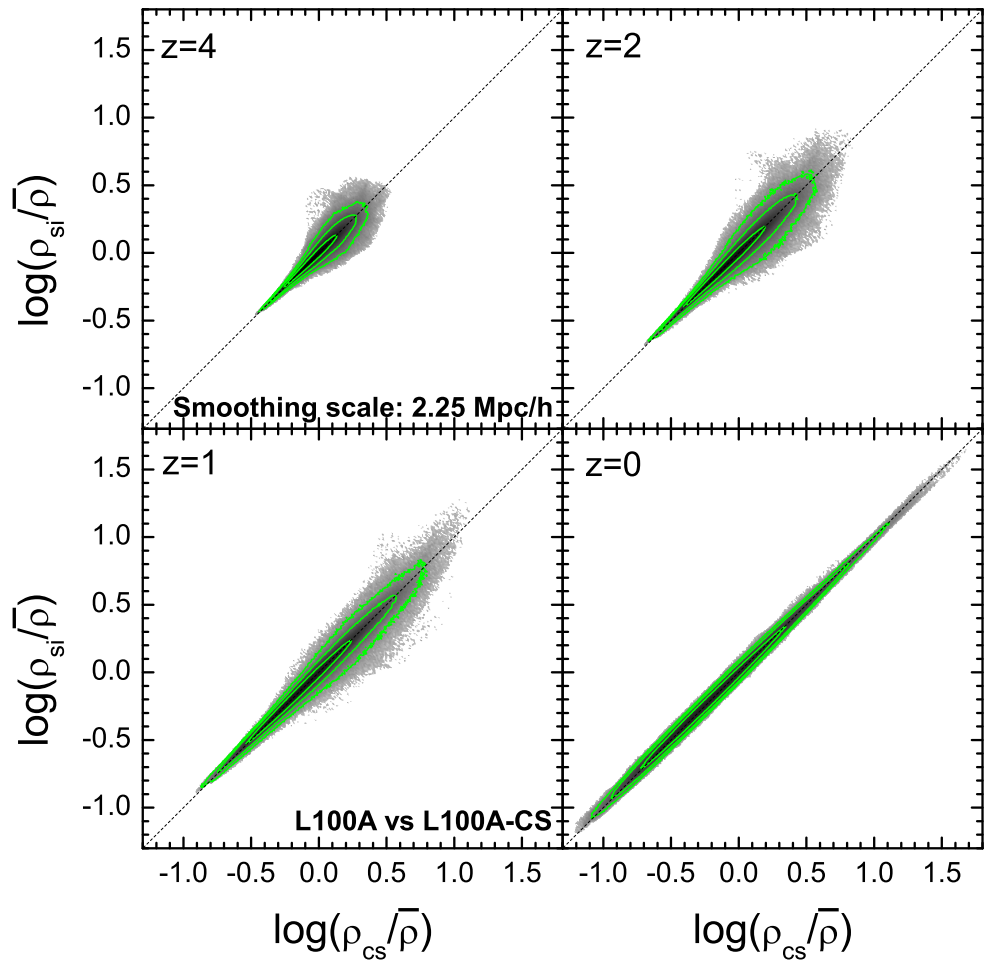,scale=1.}
\caption{The same as Figure \ref{fig_L300rho} but here for the L100A and its
CS filtered with a Gaussian function of smaller radius $2.25\mpc$.}
\label{fig_L100rho}
\end{figure*}

In Figures \ref{fig_L300rho} and \ref{fig_L100rho}, we compare
mass densities in the CSs with those in the input simulations
at four different redshifts. The density fields are smoothed within
Gaussian windows with radii the same as $R_{\rm s}$ adopted in
the corresponding HMC runs (i.e. $4.5\mpc$ for L300,
and $2.25\mpc$ for L100). The three contours in each panel
encompass 67\%, 95\% and 99\% of the grid cells in the whole simulation
box. The densities of the CS in individual cells
are tightly and linearly correlated with
the original ones over a large dynamic range; for L100 this
correlation extends from $\rho/{\bar\rho}= 0.1$ to
about $50$ at redshift zero. The constrained simulations do not show any significant bias, and at $z=0$ the typical
dispersion in the $\rho_{\rm si}$ - $\rho_{\rm cs}$ relation is about 0.05 dex
for both L300 and L100. Interestingly, the typical dispersion
does not depend significantly on redshift;  the seemingly
looser relations at higher $z$ are due to the smaller
ranges in density. However,  at high $z$ the dispersion
is larger for higher densities. The reason for this is that
small-scale modes are not well constrained in the initial
conditions and mode coupling has not fully developed
at high $z$.

\begin{figure*}
\centering
\epsfig{file=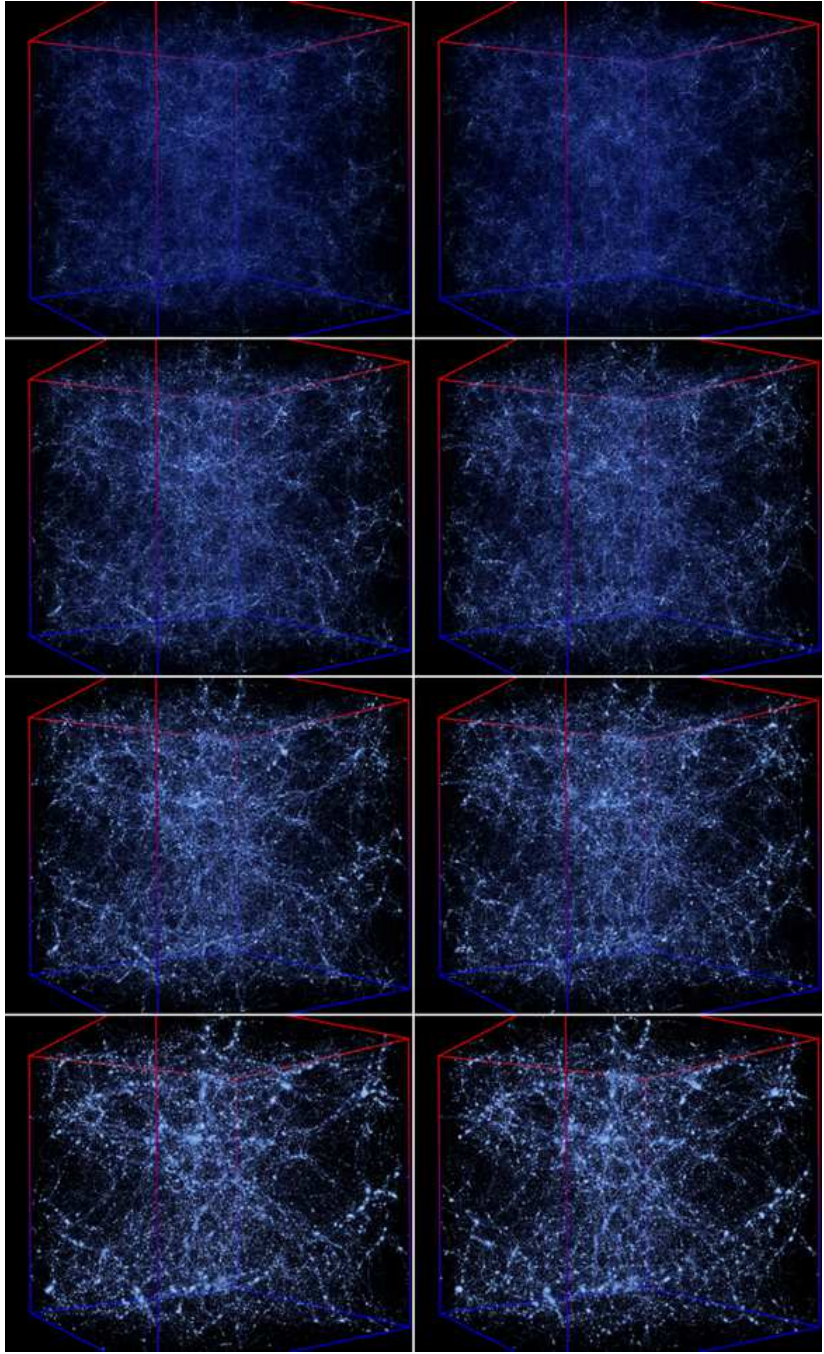,scale=1.1}
\caption{The 3-d renderings of the particle distributions, color-coded by density, in a
cubic $100\mpc$ box. The left panels show the particles in the simulation L100A
and the right panels show those in the corresponding L100A-CS. The panels
(from top to bottom) show the results at $z=4$, $2$, 1 and 0, respectively.}
\label{fig_L100A}
\end{figure*}

 Figure \ref{fig_L100A} shows some renderings of the three-dimensional
density fields of both L100A (left) and L100A-CS (right)
at four different redshifts. Particles are color-coded by local
densities calculated using the Delaunay density estimator
developed by \citet{Schaap00},
with white clumps corresponding to massive halos.
The two maps look very similar at $z=0$. Almost all large structures
in the original simulation, such as massive clusters, filaments,
and underdense voids, and even some small details, are well reproduced.
In particular, the CS accurately recovers the topological structures
of the cosmic web, suggesting that the large scale environment of
dark matter halos, which is an important factor affecting
halo formation \citep[see e.g.][]{Wang_etal11}, is well reproduced.
At higher $z$, the similarity is also remarkable, in particular
on large scales. On small scales, there are noticeable
differences which become more significant with increasing redshift,
consistent with the phase correlation results shown in
Figure \ref{fig_phasers}. These comparisons demonstrate
that our method can recover the formation history of the large-scale
structure with high accuracy.

Halos are the key components in the build-up of structure in
current CDM cosmogony and are thought to be the hosts of galaxies.
Therefore, a proper reconstruction of halo properties is essential
for our future study of galaxies and their relation with the
large-scale environment. In a forthcoming paper, we will compare
in detail the internal properties, environment and assembly histories of
halos and other structures between the input simulations and
CSs. Here we only focus on the halo mass function
$n(M_{\rm h})={\rm d}n/{\rm d}\log{M_{\rm h}}$. We identify halos
using the standard friends-of-friends (FOF) algorithm \citep{Davis_etal85}
with a link length that is $0.2$ times the mean
inter-particle separation. Following \citet{Warren_etal06}, we calculate
the mass of a halo as $M_{\rm h}=m_{\rm p}N_{\rm h}(1-N_{\rm h}^{-0.6})$,
where $m_{\rm p}$ is the particle mass and $N_{\rm h}$ is the total particle
number in a halo. This mass definition is used to correct for some
statistical effects of using only finite number of particles to
estimate the mass of a FOF halo.

\begin{figure*}
\centering
\epsfig{file=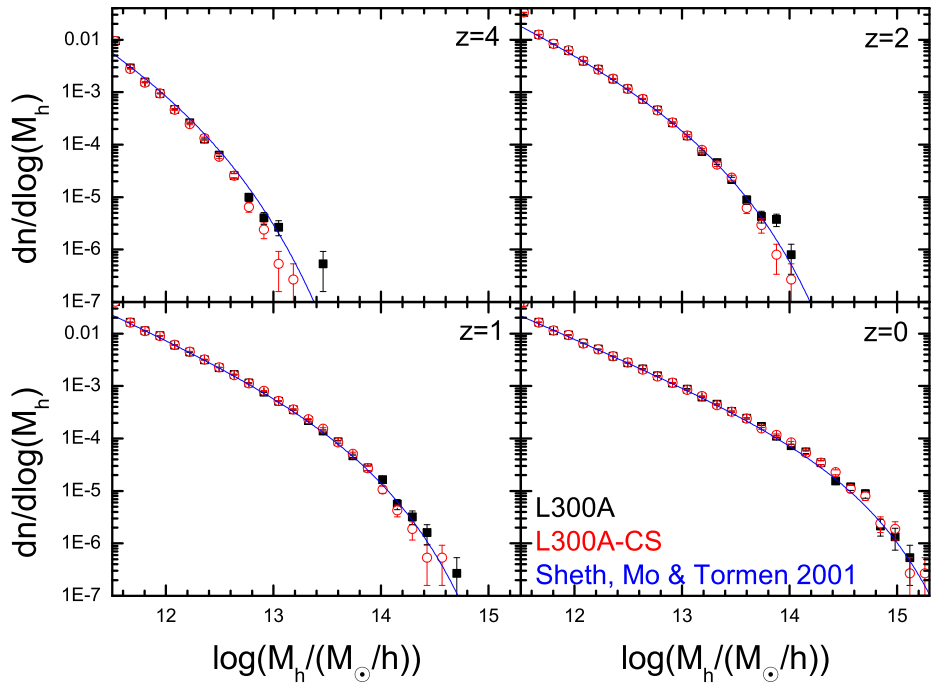,scale=1.}
\caption{Halo mass functions obtained from L300A (squares) and L300A-CS
(red circles) at four  different redshifts. The
blue lines represent the theoretical predictions \citep{Sheth_etal01}. }
\label{fig_hmfL300}
\end{figure*}

\begin{figure*}
\centering
\epsfig{file=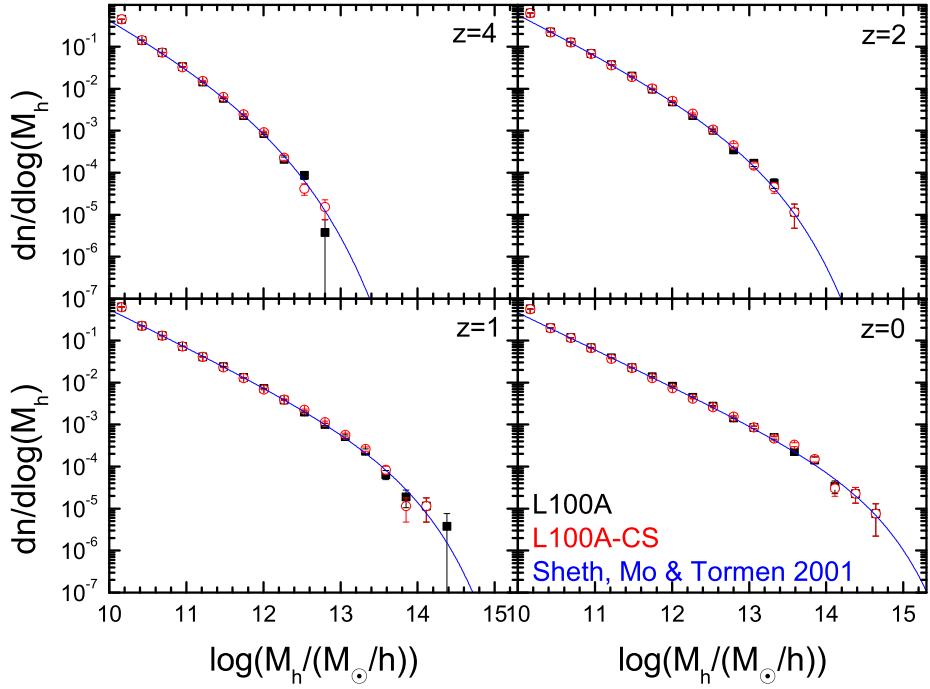,scale=1.}
\caption{The same as Figure \ref{fig_hmfL300} but here for L100A and L100A-CS.}
\label{fig_hmfL100}
\end{figure*}

\begin{figure*}
\centering
\epsfig{file=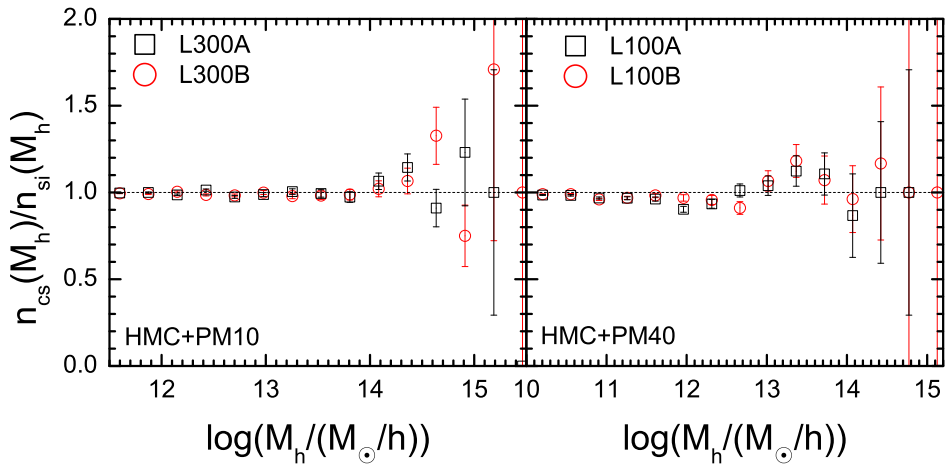,scale=1.}
\caption{The ratio between the halo mass function obtained from the
CS, $n_{\rm cs}$,  and that from the original simulation, $n_{\rm si}$.}
\label{fig_mfr1}
\end{figure*}

The halo mass functions at various redshifts obtained from
L300A-CS and L100A-CS are shown in Figures~\ref{fig_hmfL300}
and \ref{fig_hmfL100}, with the error-bars representing Poisson
fluctuations in individual mass bins. These CS mass functions are
well matched by both the theoretical prediction \citep{Sheth_etal01}
and the ones obtained from the corresponding original simulations.
To inspect the results in more detail, we show in Figure \ref{fig_mfr1}
the ratio of halo mass function between the CS and the original
simulation, $n_{\rm cs}(M_{\rm h})/n_{\rm si}(M_{\rm h})$.
If $n_{\rm si}(M_{\rm h})=0$ in a mass bin, it is replaced by
the corresponding value of the theoretical halo mass function.
In order to reduce the fluctuations at the
high mass end, halos are re-binned into fewer mass bins.
As one can see, although the halo mass
functions of the reconstructed density fields are in overall good
agreement with the original ones, there is very weak, but noticeable
bias. Such bias is not expected, as the initial density field
is well reproduced in both power spectrum and
Gaussianity. As we will see below, the reconstruction apparently
can introduce some very subtle non-Gaussian effect, which in turn
can affect the precision of the predicted halo mass function.
This effect becomes more important when a less accurate
model of structure evolution is adopted.

\section{Comparing results obtained from different models
of structure evolution}
\label{sec_dsf}

A number of different structure evolution models have been adopted
in the literature for the reconstruction of the initial cosmic density
field. Here we compare the performances of some of them, showing
that our HMC$+$PM is superior.

In our early paper (W13), we adopted the MZA model, developed by TZ12.
The phase correlation between the density field reconstructed
from the HMC$+$MAZ method with the input density field
is plotted in the upper left panel of Figure \ref{fig_phasetr}.
For this model $k_{95}\sim0.33\mpci$, very similar to
that shown in TZ12 for different cosmological parameters.
The phase correlation for 2LPT, which is adopted by \citet{JascheWandelt13} and
\citet{Kitaura13} to reconstruct the initial density fields from the
galaxy density field,  has $k_{95}=0.37\mpci$, as shown in TZ12.
More recently, \citet{Heb_etal13} employed a modification of the 2LPT,
the ALPT developed by \citet{KitauraHb13},
in their reconstruction. As shown in figure 4 of \citet{KitauraHb13},
the ALPT model can achieve $k_{95}\sim0.45\mpci$, better than the 2LPT.
For our HMC+PM method, Figure \ref{fig_Npmlc} shows
that the value of $k_{95}$ that is achieved
by PM5 and PM10 are $0.67$ and $0.80\mpci$, respectively,
much better than all these earlier models. Indeed,  our test showed that the
performances of the 2LPT and the ALPT are only between our
PM models with $N_{\rm PM}=2$ and 3.
Clearly, the PM model with a modest number of time steps
is by far the best model that can be implemented into
the HMC method.

To make further tests, linear density field is reconstructed from
the L300A simulation using both the HMC$+$MZA and the HMC$+$PM5.
The density transfer function for PM5 is given in the lower left panel
of Figure \ref{fig_phasetr}, while that for MZA is adopted from TZ12.
Since the HMC runs employing the same model of structure evolution
give very similar results (see e.g. Figure \ref{fig_phasers}),
only one run is performed for each model. Similar to the finding in W13,
the reconstructed linear power spectra from these two runs match well
the original ones, except a significant bump on intermediate scales.
We follow W13 to `renormalize' the amplitudes of the modes in the
bump so that the power spectrum matches the original one.
As shown in W13, even with the renormalization
the distributions of $d_{{\rm n},j}({\bf k})=\delta_j({\bf k})/\sqrt{P_{\rm lin}(k)/2}$
are still very close to Gaussian with $\sigma=1$. We use the
linear density fields so obtained to set up initial conditions, and follow
the evolutions of the density fields with the $N$-body code described
above.

\begin{figure*}
\centering
\epsfig{file=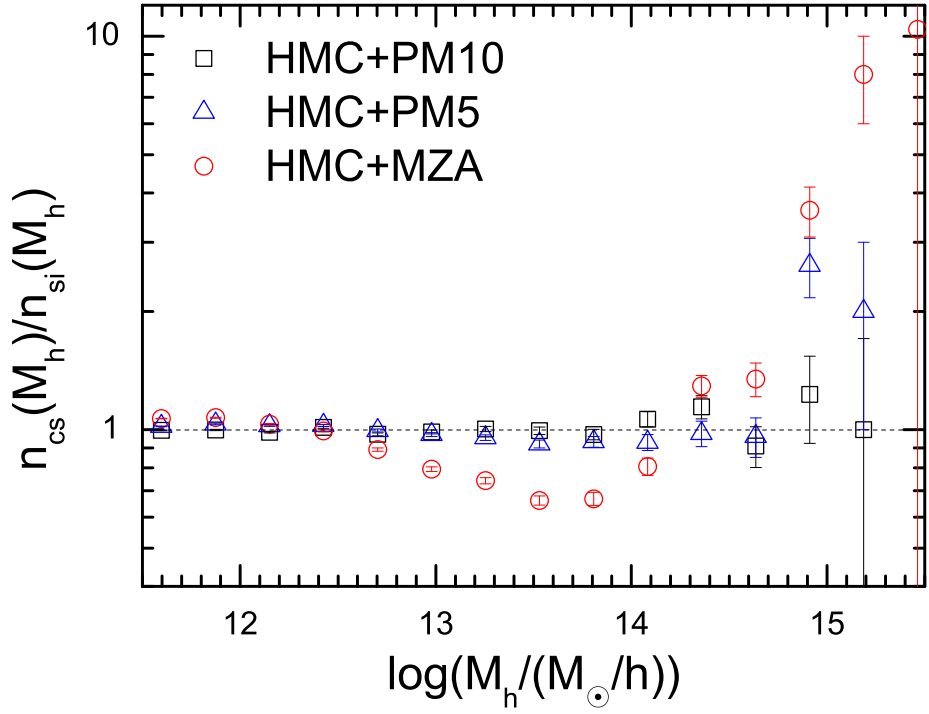,scale=1.}
\caption{The ratio between the halo mass function derived from the
reconstructed final density field and that from the original density field.
Results are shown for different models of structure evolution, as indicated in the figure.
The results are all based on L300A.}
\label{fig_mfr2}
\end{figure*}

Figure \ref{fig_mfr2} shows the mass function ratio,
$n_{\rm cs}(M_{\rm h})/n_{\rm si}(M_{\rm h})$, at $z=0$ for the
three models, HMC+PM10, HMC+PM5, and HMC+MZA.
The halo mass function for the MZA model deviates significantly
from the original one; it produces too many massive halos
at $M_{\rm h}>10^{14.3}\msun$ and too few halos of
$10^{13.5}\msun$. Second, there is a clear trend that the
mass function in the CS gradually converges to the original one
as the accuracy of the model increases, in the order from
MZA to PM5 to PM10. For PM10 the CS mass function is already quite close to the original one.
Recently, \citet{Heb_etal13} found that their CSs of the
linear density fields constructed from the ALPT model
significantly overproduce massive halos and underproduce
small-mass halos relative to the reference mass
functions. This is exactly the same as we find here when
a model of low accuracy is used to evolve the cosmic
density field, although the Bayesian approach they adopted
is different from ours. The deviation they found is therefore
more likely a result of the inaccuracy of the ALPT, rather than
due to the cosmic variance.

One possible reason for this deviation is that the HMC induces
spurious correlations in the reconstructed $\delta({\bf k})$ if the
model of structure evolution is not sufficiently accurate.
Most of the approximations, which cannot properly handle
highly non-linear dynamics on small scales, tend to underestimate
the density in high density regions, in particular around halos.
Thus the parameter $\chi^2$, which the HMC process attempts to
minimize, is contributed by two different sources.
One is the difference between the reconstructed and the real
$\delta({\bf k})$, and the other is the underestimation
of the power spectra on small scales due to the use
of approximate dynamics. During the evolution of the Hamiltonian
system, the HMC tries to tune $\delta({\bf k})$ to
compensate for the lost power in two ways. One is to enhance
the amplitude of $\delta({\bf k})$, which is seen
in the reconstructed linear power spectra
\citep[see W13;][]{Kitaura13,Ata_etal14}. The other is to introduce
some non-Gaussianity into $\delta({\bf k})$.

\begin{figure*}
\centering
\epsfig{file=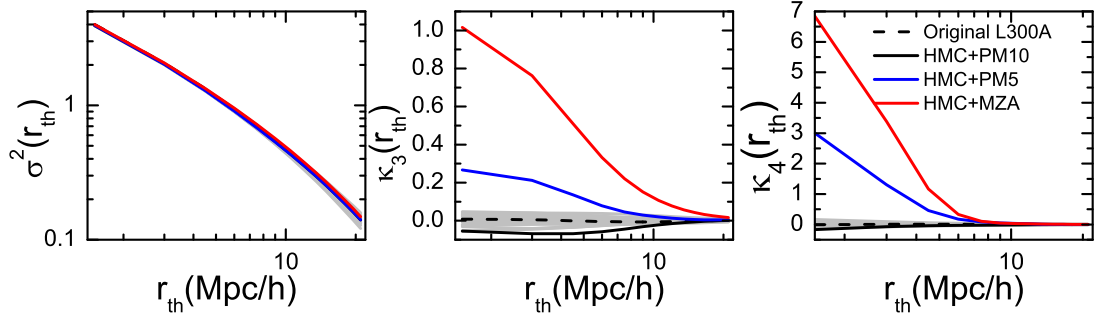,scale=1.}
\caption{The variance ($\sigma^2$), skewness ($\kappa_{3}$) and kurtosis ($\kappa_{4}$) of the linear density fields in real space as a function of smoothing scale $r_{\rm th}$ (top-hat smoothing kernel). The black dashed lines show the results measured from the original linear density field of L300A. The solid lines show the results measured from the reconstructed linear density fields based on different models of
structure evolution, as indicated in the right panel. The grey bands show the spread
of 19 linear density fields randomly sampled from the prior Gaussian distribution,
$G(\delta({\bf k}))$, given in equation (\ref{eq_post}).}
\label{fig_k3k4}
\end{figure*}

To verify the latter probability, we first derive a smoothed version
of the reconstructed linear density field, $\delta({\bf x}, r_{\rm th})$,
where the smoothing is done with a top-hat kernel
with radius $r_{\rm th}$. We then measure the variance
($\kappa_2\equiv\sigma^2$), skewness ($\kappa_3$) and
kurtosis ($\kappa_4$) of the smoothed field. For a random
Gaussian field, it is easy to prove that $\kappa_l=0$ for all $l>2$.
Thus any non-zero high-order $\kappa_l$ signifies
non-Gaussianity. Figure \ref{fig_k3k4} shows the three quantities
as functions of $r_{\rm th}$ for the original linear density field and
for the reconstructed linear density fields based on PM10,
PM5 and MZA. For comparison, we also present the results
measured from 19 random Gaussian fields.

The three reconstructions exhibit almost identical $\sigma^2(r_{\rm th})$
as the original one, which is expected since $\sigma^2(r_{\rm th})$ is directly
related to the linear power spectrum.
However there are clear differences in $\kappa_3$ and $\kappa_4$
between the three reconstructions and the original linear density field.
By construction, these two quantities should be close to zero. The reconstructed linear density
field based on PM10 exhibits small nonzero $\kappa_3$ on small
scales, but the deviation is still  within the grey region which represents
the variance among the 19 random Gaussian fields.
The situation is very similar for $\kappa_4$. Thus, no significant
signal for non-Gaussianity is present in the reconstructed
$\delta({\bf k})$ based on PM10. In contrast, the
non-Gaussianity is quite significant for PM5 and is the strongest
for MZA. Such non-Gaussianity is hard to see in the distribution
function of $\delta({\bf k})$ but is the source of bias in
the predicted halo mass function. As the densest and massive objects
in the cosmic density field, massive halos are sensitive
to the tail of the density distribution. These tests support our
hypothesis that the origin of the deviation in halo mass function
is non-Gaussianity.

Another potential issue is the smoothing scale $R_{\rm s}$
adopted for the three tests shown here, all being
$3l_{\rm c}=4.5\mpc$.  According to our results of the two-point
correlation function analysis (Section \ref{sec_tpm}), this smoothing scale
can effectively erase the difference between the model prediction
of PM10 and the simulation, and so the result of the halo
mass function is not biased by the inaccuracy of the model.
For PM5 and MZA, however, $R_{\rm s}=4.5\mpc$ may
be too small to get rid of the difference between the model
and the simulation. This again shows
that, in order  to decrease the smoothing scale $R_{\rm s}$,
so as to increase the accuracy of the reconstruction,
and simultaneously to maintain Gaussianity, a highly
accurate model of structure evolution is required.
For example, the HMC+PM40 model with $R_{\rm s}=2.25\mpc$
predict a phase correlation that is significantly tighter
than the prediction of the HMC+PM10 model.
Consequently the halo mass function obtained is
also in better agreement with the original one.

\section{Summary and Discussion}
\label{sec_sum}

Following the spirit of our previous paper \citep[W13, see also][]{JascheWandelt13},
we have developed a HMC method to reconstruct the
linear (initial) density field from a given input non-linear density
field at low redshift. This method allows us to generate the linear
density field based on a posterior probability distribution function
including both a prior term and a likelihood term. The prior term
ensures that the reconstructed linear density field obeys a Gaussian
distribution with a given power spectrum. The likelihood is designed
to ensure that the evolved density field from the reconstructed
linear density field matches the input non-linear density field.
To improve the accuracy of the reconstruction, we have combined
the HMC with the PM dynamics. The Hamiltonian force is the most
important quantity that guides the chain steps,  and we have
worked out how to compute the Hamiltonian force for the PM model.

To optimize the parameters in the PM model, we have made
a series of tests to investigate how the quality of the PM model
predictions depend on the grid cell size, $l_{\rm c}$,  and
the number of time steps, $N_{\rm PM}$. The minimum $N_{\rm PM}$
required for the PM model to approach its best quality
roughly scales as $1/l^2_{\rm c}$, and the wavenumber below
which the PM model is reliable scales as $1/l_{\rm c}$.
As a compromise between the accuracy of model prediction
and computation time,  we have made HMC runs with two
choices, one with $l_{\rm c}=1.5\mpc$, and the corresponding
minimum number of timesteps is $N_{\rm PM}=10$; the other
one has $l_{\rm c}=0.75\mpc$ and the minimum number of
time steps is $N_{\rm PM}=40$. To increase the accuracy of model
prediction, a density transfer function, which is very easy to obtain,
is used.  The PM10 model can produce more than 95\% of the
phase information in the corresponding high-resolution
$N$-body simulation down to a scale
corresponding to $k=0.8\mpci$. The PM40 model
is much more accurate,  recovering more than 95\% of the
phase information all the way down to a highly non-linear
scale, $k=1.78\mpci$. The discrepancy between the
PM prediction and the input field on small scales need to
be suppressed in a HMC run. Based on our two-point
correlation function analysis, this can be done with
a Gaussian filter with a radius $R_{\rm s}\approx  3l_{\rm c}$.

We have used four high-resolution $N$-body simulations as inputs
to test the performance of our HMC+PM method.
The Fourier modes of the reconstructed linear density fields obey
well a Gaussian distribution with variance that is well matched by
the original linear power spectrum.The phase-correlation coefficient
between the reconstructed and original linear density fields
is close to unity on large scales and declines gradually to 0.5 at
$k\sim0.85\mpci$ and $k\sim0.47\mpci$ for the HMC+PM40
and HMC+PM10 runs, respectively. The {\it RMS} in the difference between
the original (input) and modeled density fields are only about 4.6\%.
A weak discrepancy appears around the scale where the ratio
between the likelihood and prior terms of the Hamiltonian force
is about one, and is due to the compromise the HMC makes
between these two terms.

As additional tests, we have compared the original simulations
with the corresponding constrained simulations (CS) that are evolved
from the reconstructed linear density fields with a high-resolution
$N$-body code. The density fields in the CS are found to match the original
ones over a large dynamic range at various redshifts. The typical
dispersions in the density-density relation are about 0.05 dex on the scale
of $3l_{\rm c}$, almost independent of redshift.
Visual inspection of the particle distributions also shows that the
CSs can well reproduce the large scale structures and their evolution histories.
For the HMC+PM40 (HMC+PM10) run, the phase correlation between
the CS and the original simulation is very close to unity down to
a scale corresponding to $k=1.1\mpci$ ($0.5\mpci$),
and remains larger than 0.5 all the way to the highly non-linear
regime $k\sim3.4\mpci$ ($1.1\mpci$). In addition the halo mass
functions derived from the CSs match well those derived from
the original simulations. The tests based on the four different
simulations give very similar results, demonstrating the robustness
of our method.

We have also investigated how the performance of the HMC method depends
on the accuracy of the dynamical model of structure evolution adopted.
For a given $R_{\rm s}$, a low-quality dynamical model results in a
large deviation in the halo mass function, and produces
significant non-Gaussianity in the reconstructed linear density
field. All these results demonstrate the importance of
adopting accurate dynamics in the HMC approach of reconstruction,
and our HMC$+$PM method provides such a scheme. The method can achieve very high accuracy as long as the available computational
resource allows the implementation of a PM model with sufficiently
small $l_{\rm c}$ and sufficiently large $N_{\rm PM}$.

In the future, we will apply our HMC+PM method to generate the initial conditions for the structure formation in the local Universe using observational data. Since our method needs an input density field, the method developed
by \citet{Wang_etal09} will be adopted to reconstruct the
present-day density field from the SDSS DR7 galaxy group catalog constructed by \citet{Yang_etal07}. This halo-based reconstruction
method has been tested in great detail in Wang et al. (2009; 2012; 2013) and has been applied to SDSS DR4 by \citet{Munoz-Cuartas_etal11} and to SDSS DR7 by W13. Our HMC+PM method can also be applied to the $z\sim2$ density field reconstructed from the Ly$\alpha$ forest data \citep{Pichon_etal01,Caucci_etal08}. The test based on mock spectra by \citet{Lee_etal14} suggested that the spatial resolution in such reconstruction can reach to a scale of $\sim2\mpc$ (in comoving units) at $z\sim2$,
which can be handled very easily with our method. These studies showed that there are significant uncertainties in the
reconstructed density fields at a given redshift, which may eventually
be the main source of uncertainties in future applications.
We will come back to this in a forthcoming paper.

The reconstructed initial conditions can be used to run CSs
with both dark matter and gas components. This
offers a unique opportunity to investigate the formation and evolution
of the galaxy population we directly observe. For example, we
can model galaxy formation using halos extracted from the
CSs combined with semi-analytical models of galaxy formation,
or carry out gas simulations directly from the reconstructed
linear density field. The comparison between the modeled
galaxy population and  the observed one
is then conditioned on the same large-scale
environments, and so the effect of cosmic variance is eliminated or
reduced. Furthermore, the predicted distribution and state of the
gas component can also be compared directly with observations
of the Sunyaev-Zel'dovich effect, X-ray emission, and quasar
absorption line systems, to provide a more complete picture
about the interaction between dark matter, gas and  galaxies.

\section*{Acknowledgments}

We thank Yu Yu for useful suggestion and Kun Cheng for rendering
the figures. This work is supported by the Strategic Priority Research Program "The Emergence of Cosmological Structures" of the Chinese Academy of Sciences, Grant No. XDB09010400, NSFC (11073017, 11233005, 11320101002, 11121062, 11033006, U1331201), NCET-11-0879. HJM would like to acknowledge the support of NSF AST-1109354 and the CAS/SAFEA International Partnership Program for Creative  Research Teams (KJCX2-YW-T23).

\appendix

\section{Abbreviations, Terms and Quantities}\label{app_abb}

\begin{itemize}
\item  PM model:           Particle-Mesh model;
\item $l_{\rm c}$:           the mesh cell size in a PM model;
\item $N_{\rm PM}$:      number of time-steps in a PM model;
\item PM$N_{\rm PM}$:  a PM model with $N_{\rm PM}$ time-steps;
\item PM$N_{\rm PM}$ density field:
               the density field predicted by the corresponding PM model;
\item Modeled density field: the density field predicted by a dynamical model of structure evolution;
\item MZA:     modified Zel'dovich approximation;
\item HMC:     Hamiltonian Markov Chain Monte Carlo;
\item HMC$+$PM$N_{\rm PM}$:  a HMC run using a specific PM model;
\item HMC$+$MZA:                      a HMC run using MZA;
\item Input density field or $\rho_{\rm inp}$:
                                   the density field used to constrain a reconstruction;
\item Original quantities: quantities measured from or related to an input simulation;
\item $R_{\rm s}$: the radius of the Gaussian kernel used to smooth both the
                    input density field and the modeled density
                    fields in a HMC run;
\item Reconstructed linear density field: the linear density field
                    reconstructed from an input density field;
\item Reconstructed linear power spectrum: the linear power spectrum measured
                     from a reconstructed linear density field;
\item CS: constrained simulation, the simulation of a reconstructed
                    linear density field;
\item CS quantities: quantities measured from a CS.
\end{itemize}

We test HMC with three PM models: PM5, PM10 and PM40. The mesh cell size
($l_{\rm c}$) for PM5 and PM10 is $1.5\mpc$, and is $0.75\mpc$ for PM40.

\section{Calculations of the likelihood term of Hamiltonian force}
\label{app_hf}

Here, we provide the details of the Hamiltonian force
calculation described in Section \ref{sec_hf}. For conciseness, we use $N$ to replace $N_{\rm PM}$ in this appendix.

\subsection{The $\chi^2$ transformation}
\label{sec_hf1}

Following equation (13) in W13, we have
\begin{equation}\label{eq_pi1}
\frac{\partial\chi^2}{\partial {\bf r}_{N}({\bf q})}
=N_{\rm c}^3\sum_{\bf
k}\rho_{\rm d}^{\ast}({\bf k})\frac{\partial\rho_{\rm mod}({\bf
k})}{\partial{\bf r}_{N}({\bf q})}\,,
\end{equation}
where $\rho_{\rm d}({\bf x})=[\rho_{\rm mod}({\bf
x})-\rho_{\rm inp}({\bf x})]\omega({\bf x})/\sigma_{\rm inp}^2({\bf x})$.
Inserting the expression of $\rho_{\rm mod}$ given by
equation (\ref{eq_rhomod}) into the above expression, we have
\begin{eqnarray}\label{eq_pi2}
\frac{\partial\chi^2}{\partial {\bf r}_{N}({\bf q})}&=&\sum_{{\bf x}}\frac{\partial\rho_{N,\rm c}({\bf x})}{\partial{\bf r}_N({\bf q})}\sum_{\bf k}\rho^{\ast}_{\rm d}({\bf k})\frac{w_{\rm g}(R_{\rm s}k)T(k)}{w_{\rm c}({\bf k})}{\rm e}^{i{\bf k}\cdot({\bf L}-{\bf x})}\nonumber\\
&=&\sum_{{\bf x}}\sum_{{\bf q}_1}\frac{\partial w_{\rm c}({\bf x}-{\bf r}_N({\bf q}_1))}{\partial{\bf r}_N({\bf q})}\rho_{\rm ds}({\bf L}-{\bf x})=\sum_{{\bf x}}\frac{\partial w_{\rm c}({\bf x}-{\bf r}_N({\bf q}))}{\partial{\bf r}_N({\bf q})}\rho_{\rm ds}({\bf L}-{\bf x})\,,
\end{eqnarray}
where $\rho_{\rm ds}({\bf x})$ is the Fourier transformation of
$\rho^{\ast}_{\rm d}({\bf k})w_{\rm g}(R_{\rm s}k)T(k)/w_{\rm c}({\bf k})$ and ${\bf L}=(L,L,L)$ so ${\rm e}^{i{\bf k}\cdot\bf L}\equiv 1$. Note that $\partial
w_{\rm c}({\bf x}-{\bf r}_N({\bf q}_1))/\partial{\bf r}_N({\bf q})$ is
nonzero only when ${\bf q}_1=\bf q$. Moreover, $\partial w_{\rm
 c}({\bf x}-{\bf r}_N({\bf q}))/\partial{\bf r}_N({\bf q})$ is
nonzero only in a number of grid cells close to ${\bf r}_N({\bf q})$,
which can also be used to speed up the calculation of equation
(\ref{eq_pi2}). Note that $\chi^2$ is independent of the final
velocity field so that ${\partial\chi^2}/{\partial {\bf v}_{N-1/2}({\bf q})}=0$.

\subsection{The Particle-Mesh transformation}
\label{sec_hf2}

Starting from the last step $n=N$ where the derivatives of $\chi$
are obtained, we can obtain the derivatives successively at other
steps. Suppose we have already obtained
$\partial\chi^2/\partial {\bf r}_{n+1}({\bf q})$
and $\partial\chi^2/\partial {\bf v}_{n+1/2}({\bf q})$.
We can use the chain rule to write
$\partial\chi^2/\partial {\bf r}_{n}({\bf q})$ and
$\partial\chi^2/\partial {\bf v}_{n-1/2}({\bf q})$ as
\begin{equation}\label{eq_pmt1}
\frac{\partial\chi^2}{\partial {\bf v}_{n-1/2}({\bf q})}
=\sum_{{\bf q}_1}\left[\frac{\partial{\bf r}_{n+1}({\bf q}_1)}{\partial{\bf v}_{n-1/2}({\bf q})}\otimes\frac{\partial\chi^2}{\partial {\bf r}_{n+1}({\bf q}_1)}+\frac{\partial{\bf v}_{n+1/2}({\bf q}_1)}{\partial{\bf v}_{n-1/2}({\bf q})}\otimes\frac{\partial\chi^2}{\partial {\bf v}_{n+1/2}({\bf q}_1)}\right]\,,
\end{equation}
\begin{equation}\label{eq_pmt2}
\frac{\partial\chi^2}{\partial {\bf r}_{n}({\bf q})}
=\sum_{{\bf q}_1}\left[\frac{\partial{\bf r}_{n+1}({\bf q}_1)}{\partial{\bf r}_{n}({\bf q})}\otimes\frac{\partial\chi^2}{\partial {\bf r}_{n+1}({\bf q}_1)}+\frac{\partial{\bf v}_{n+1/2}({\bf q}_1)}{\partial{\bf r}_{n}({\bf q})}\otimes\frac{\partial\chi^2}{\partial {\bf v}_{n+1/2}({\bf q}_1)}\right]\,,
\end{equation}
where the symbol $\otimes$ stands for matrix
multiplication. According to the leapfrog equations
(\ref{eq_pm6}) and (\ref{eq_pm7}) in the PM model, we have that
\begin{equation}
\frac{\partial{\bf r}_{n+1}({\bf q}_1)}{\partial{\bf r}_{n}({\bf q})}={\bf I}\delta({{\bf q}_1-{\bf q}})+\frac{\partial{\bf F}_n({\bf r}_n({\bf q}_1))}{\partial{\bf r}_n({\bf q})}\Delta^{\rm v}_n\Delta^{\rm r}_n\;,
\end{equation}
\begin{equation}
\frac{\partial{\bf r}_{n+1}({\bf q}_1)}{\partial{\bf v}_{n-1/2}({\bf q})}={\bf I}\delta({{\bf q}_1-{\bf q}})\Delta^{\rm r}_n\;,
\end{equation}
\begin{equation}
\frac{\partial{\bf v}_{n+1/2}({\bf q}_1)}{\partial{\bf r}_{n}({\bf q})}=\frac{\partial{\bf F}_n({\bf r}_n({\bf q}_1))}{\partial{\bf r}_n({\bf q})}\Delta^{\rm v}_n\;,
\end{equation}
\begin{equation}
\frac{\partial{\bf v}_{n+1/2}({\bf q}_1)}{\partial{\bf v}_{n-1/2}({\bf q})}={\bf I}\delta({{\bf q}_1-{\bf q}})\;,
\end{equation}
where $\bf I$ is a unit matrix, and $\delta({{\bf q}_1-{\bf q}})$
is equal to one when ${\bf q}_1={\bf q}$, and zero otherwise.
In the derivation, we have used the facts that
$\partial{\bf r}_{n}({\bf q}_1)/\partial{\bf r}_{n}({\bf
  q})=\delta({\bf q}_1-{\bf q})\bf I$ and $\partial{\bf r}_{n}({\bf
  q}_1)/\partial{\bf v}_{n-1/2}({\bf q})=0$

Inserting these equations into equations (\ref{eq_pmt1}) and (\ref{eq_pmt2}), we obtain
\begin{equation}\label{eq_pmt3}
\frac{\partial\chi^2}{\partial {\bf v}_{n-1/2}({\bf q})}=\frac{\partial\chi^2}{\partial {\bf v}_{n+1/2}({\bf q})}+\frac{\partial\chi^2}{\partial {\bf r}_{n+1}({\bf q})}\Delta^{\rm r}_n\,,
\end{equation}
\begin{equation}\label{eq_pmt4}
\frac{\partial\chi^2}{\partial {\bf r}_{n}({\bf q})}=\frac{\partial\chi^2}{\partial {\bf r}_{n+1}({\bf q})}+\Delta^{\rm v}_n\sum_{{\bf q}_1}\frac{\partial{\bf F}_n({\bf r}_n({\bf q}_1))}{\partial{\bf r}_n({\bf q})}\otimes\frac{\partial\chi^2}{\partial {\bf v}_{n-1/2}({\bf q}_1)}=\frac{\partial\chi^2}{\partial {\bf r}_{n+1}({\bf q})}+\Delta^{\rm v}_n{\bf F}^{\chi}_{n}({\bf q})\,.
\end{equation}
Using equation (\ref{eq_pmforce}), the quantity
${\bf F}^{\chi}_{n}({\bf q})$ defined in the second term can be written as
\begin{eqnarray}\label{eq_pmt5}
{\bf F}^{\chi}_{n}({\bf q})&\equiv&\sum_{{\bf q}_1}\frac{\partial{\bf F}_n({\bf r}_n({\bf q}_1))}{\partial{\bf r}_n({\bf q})}\otimes\frac{\partial\chi^2}{\partial {\bf v}_{n-1/2}({\bf q}_1)}\nonumber\\
&=&\sum_{{\bf q}_1}\sum_{{\bf x}_1}w_{\rm c}({\bf x}_1-{\bf r}_n({\bf q}_1))\frac{\partial{\bf F}_{{\rm g},n}({\bf x}_1)}{\partial{
\bf r}_n({\bf q})}\otimes\frac{\partial\chi^2}{\partial {\bf v}_{n-1/2}({\bf q}_1)}\nonumber\\
&+&\sum_{{\bf q}_1}\sum_{{\bf x}_1}\frac{\partial w_{\rm c}({\bf x}_1-{\bf r}_n({\bf q}_1))}{\partial{\bf r}_n({\bf q})}\left[{\bf F}_{{\rm g},n}({\bf x}_1)\cdot\frac{\partial\chi^2}{\partial {\bf v}_{n-1/2}({\bf q}_1)}\right]\nonumber\\
&=&{\bf F}^{\chi}_{n,a}({\bf q})+{\bf F}^{\chi}_{n,b}({\bf q})\,,
\end{eqnarray}
where `$\cdot$' stands for dot product, and the last line of
the equation defines  ${\bf F}^{\chi}_{n,a}({\bf q})$ and
${\bf F}^{\chi}_{n,b}({\bf q})$. The first term of
${\bf F}^{\chi}_{n}({\bf q})$ can be rewritten as
\begin{eqnarray}\label{eq_pmt6}
{\bf F}^{\chi}_{n,a}({\bf q})&\equiv &\sum_{{\bf q}_1}\sum_{{\bf x}_1}w_{\rm c}({\bf x}_1-{\bf r}_n({\bf q}_1))\frac{\partial{\bf F}_{{\rm g},n}({\bf x}_1)}{\partial{
\bf r}_n({\bf q})}\otimes\frac{\partial\chi^2}{\partial {\bf v}_{n-1/2}({\bf q}_1)}\nonumber\\
&=&\sum_{{\bf x}_2}\frac{\partial\delta_{n, \rm c}({\bf x}_2)}{\partial{\bf r}_n({\bf q})}\sum_{\bf k}\frac{w_{\rm g}(R_{\rm PM}k)}{w^2_{\rm c}({\bf k})N_c^3}{\rm e}^{-i{\bf k}\cdot{\bf x}_2}\frac{(i\bf k\cdot)}{k^2}\sum_{{\bf x}_1}{\rm e}^{i{\bf k}\cdot{\bf x}_1}\sum_{{\bf q}_1}w_{\rm c}({\bf x}_1-{\bf r}_n({\bf q}_1))\frac{\partial\chi^2}{\partial{\bf v}_{n-1/2}({\bf q}_1)}\nonumber\\
&=&\sum_{{\bf x}_2}\sum_{{\bf q}_2}\frac{\partial w_{\rm c}({\bf x}_2-{\bf r}_n({\bf q}_2))}{\partial{\bf r}_n({\bf q})}\sum_{\bf k}\frac{w_{\rm g}(R_{\rm PM}k)}{w^2_{\rm c}({\bf k})N_c^3}{\rm e}^{-i{\bf k}\cdot{\bf x}_2}\frac{(i\bf k\cdot)}{k^2}\sum_{{\bf x}_1}{\rm e}^{i{\bf k}\cdot{\bf x}_1}{\bf d}_{v,n}({\bf x}_1)\nonumber\\
&=&\sum_{{\bf x}_2}\frac{\partial w_{\rm c}({\bf x}_2-{\bf r}_n({\bf q}))}{\partial{\bf r}_n({\bf q})}\sum_{\bf k}{\rm e}^{i{\bf k}\cdot({\bf L}-{\bf x}_2)}\frac{w_{\rm g}(R_{\rm PM}k)}{w^2_{\rm c}({\bf k})}\frac{(i{\bf k}\cdot{\bf d}^{\ast}_{v,n}({\bf k}))}{k^2}\nonumber\\
&=&\sum_{{\bf x}_2}\frac{\partial w_{\rm c}({\bf x}_2-{\bf r}_n({\bf q}))}{\partial{\bf r}_n({\bf q})}f_{v,n}({\bf L}-{\bf x}_2)\,,
\end{eqnarray}
where $f_{v,n}({\bf L}-{\bf x}_2)$ is defined by the last line of the
equation, and
\begin{equation}
{\bf d}_{v,n}({\bf x}_1)\equiv
\sum_{{\bf q}_1}w_{\rm c}({\bf x}_1-{\bf r}_n({\bf q}_1))
\frac{\partial\chi^2}{\partial{\bf v}_{n-1/2}({\bf q}_1)}\,.
\end{equation}
Thus, ${\bf d}_{v,n}({\bf x}_1)$ is equivalent to CIC assignments
of `particles' located at ${\bf r}_n({\bf q}_1)$ with
`masses'  given by $\partial\chi^2/\partial{\bf v}_{n-1/2}({\bf q}_1)$
to grid points. Its Fourier transform, ${\bf d}_{v,n}({\bf k})$,
can be obtained readily. Since
$\partial w_{\rm c}({\bf x}_2-{\bf r}_n({\bf q}))/\partial{\bf
  r}_n({\bf q})$ is nonzero only at a small number of grid
cells close to ${\bf r}_n({\bf q})$, the calculation implied by
the last line of equation (\ref{eq_pmt6}) is not time-consuming.
The second term of ${\bf F}^{\chi}_{n}({\bf q})$ can be
rewritten as
\begin{eqnarray}\label{eq_pmt7}
{\bf F}^{\chi}_{n,b}({\bf q})&=&\sum_{{\bf q}_1}\sum_{{\bf x}_1}\frac{\partial w_{\rm c}({\bf x}_1-{\bf r}_n({\bf q}_1))}{\partial{\bf r}_n({\bf q})}\left[{\bf F}_{{\rm g},n}({\bf x}_1)\cdot\frac{\partial\chi^2}{\partial {\bf v}_{n-1/2}({\bf q}_1)}\right]\nonumber\\
&=&\sum_{{\bf x}_1}\frac{\partial w_{\rm c}({\bf x}_1-{\bf r}_n({\bf q}))}{\partial{\bf r}_n({\bf q})}\left[{\bf F}_{{\rm g},n}({\bf x}_1)\cdot\frac{\partial\chi^2}{\partial {\bf v}_{n-1/2}({\bf q})}\right]\,.
\end{eqnarray}

The above transformation can be carried out until $n=0$ to give
$\partial\chi^2/\partial {\bf r}_{\rm i}({\bf q})=
\partial\chi^2/\partial {\bf r}_0({\bf q})$ and
$\partial\chi^2/\partial {\bf v}_{\rm i}({\bf q})=
\partial\chi^2/\partial {\bf v}_0({\bf q})$.

\subsection{The Zel'dovich transformation}
\label{sec_hf3}

Finally, the likelihood term of the Hamiltonian force can be written as
\begin{equation}
F_{j}({\bf k})=\frac{\partial\chi^2}{\partial\delta_j({\bf k})}=\sum_{\bf q}\left[\frac{\partial\chi^2}{\partial{\bf r}_{\rm i}({\bf q})}\cdot\frac{\partial{\bf r}_{\rm i}({\bf q})}{\partial\delta_j({\bf k})}+\frac{\partial\chi^2}{\partial{\bf v}_{\rm i}({\bf q})}\cdot\frac{\partial{\bf v}_{\rm i}({\bf q})}{\partial\delta_j({\bf k})}\right]\,,
\end{equation}
Inserting equations (\ref{eq_za1}) and (\ref{eq_za2}) into the above equation, we have
\begin{equation}
F_{j}({\bf k})=\sum_{\bf q}\frac{\partial{\bf s}({\bf q})}{\partial\delta_j({\bf k})}\cdot\left[\frac{\partial\chi^2}{\partial{\bf r}_{\rm i}({\bf q})}+\frac{\partial\chi^2}{\partial{\bf v}_{\rm i}({\bf q})}H_{\rm i}a^2_{\rm i}f(\Omega_{\rm i})\right]=\frac{1}{N_{\rm c}^3}\sum_{\bf q}\frac{\partial{\bf s}({\bf q})}{\partial\delta_j({\bf k})}\cdot{\bf \Psi}({\bf q})\,.
\end{equation}
With the use of the Fourier transform of ${\bf s}({\bf q})$
given in equation (\ref{eq_za3}), the Hamiltonian force can be rewritten as,
\begin{equation}
F_{j}({\bf k})=\sum_{{\bf k}_1}\frac{D(a_{\rm i})}{k^2_1}\frac{\partial\delta({\bf k}_1)}{\partial\delta_j({\bf k})}(i{\bf k}_1\cdot)\frac{1}{N_{\rm c}^3}\sum_{\bf q}{\bf \Psi}({\bf q}){\rm e}^{i{\bf k}_1\cdot\bf q}=\sum_{{\bf k}_1}\frac{D(a_{\rm i})}{k^2_1}\frac{\partial\delta({\bf k}_1)}{\partial\delta_j({\bf k})}(i{\bf k}_1\cdot{\bf \Psi}^{\ast}({\bf k}_1))\,,
\end{equation}
where ${\bf \Psi}({\bf k}_1)$ is the Fourier transform of ${\bf \Psi}({\bf q})$.
Since $\partial\delta({\bf k}_1)/\partial\delta_j({\bf k})$ is
nonzero only when ${\bf k}_1=\pm{\bf k}$, we eventually obtain
the expression of the likelihood term of the
Hamiltonian force for the real part of $\delta({\bf k})$ as
\begin{equation}\label{eq_af0}
F_{\rm re}({\bf k})=\frac{2D(a_{\rm i})}{k^2}{{\bf k}\cdot{\bf\Psi}_{\rm im}({\bf k})}\,,
\end{equation}
and for the imaginary part as
\begin{equation}\label{eq_af1}
F_{\rm im}({\bf k})=-\frac{2D(a_{\rm i})}{k^2}{{\bf k}\cdot{\bf\Psi}_{\rm re}({\bf k})}\,,
\end{equation}
where ${\bf\Psi}_{\rm re}({\bf k})$ and ${\bf\Psi}_{\rm im}({\bf k})$
are the real and imaginary parts of ${\bf\Psi(k)}$, respectively.

\section{Test of the HMC$+$PM method using PM
density fields as input}
\label{app_apd}

As shown in Section \ref{sec_asd}, there is always a small bump in the
reconstructed linear power spectrum relative to the original one.
There are two possibilities for this bias. One is the inaccuracy of
the adopted dynamical model, and the other is due to the HMC
method itself.  In order to distinguish these two possibilities, we
apply the HMC method to density fields generated by the PM model,
so that the first possibility is not an issue and any remaining bias
should be due to the HMC method.  Two PM density fields in
periodic boxes of 300 and $100\mpc$ are used for the test.
The density field in the larger box is generated by using PM10,
while the smaller box by using PM40. When applying
our method to these PM density fields, exactly the same PM models
are implemented in the HMC runs. Thus, the PM models adopted
in these HMC runs can be regarded as 100 percent accurate.

\begin{figure*}
\centering
\epsfig{file=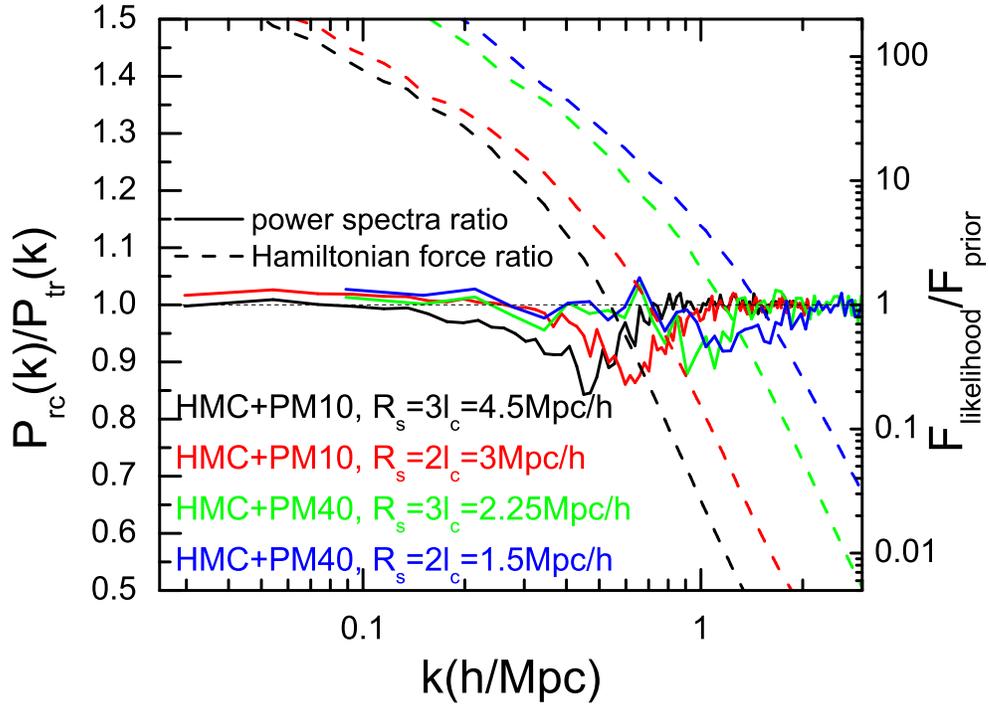,scale=1.}
\caption{The solid lines show the reconstructed linear power spectra, normalized by
the true power spectrum (left axis), $P_{\rm rc}/P_{\rm tr}$.
The dashed lines show the ratios between the likelihood term and prior term
of the Hamiltonian force (right axis). These reconstructions are for
PM density fields as input, as detailed in Appendix \ref{app_apd}.}
\label{fig_psr}
\end{figure*}

Four HMC tests are performed . Two are applied to the PM10 density
fields smoothed on scales of $R_{\rm s}=4.5\mpc$ and $3\mpc$,
respectively. The other two use the PM40 density fields smoothed with
$R_{\rm s}=2.25\mpc$ and $1.5\mpc$ as inputs. Figure \ref{fig_psr} shows
the power spectrum ratio, $P_{\rm rc}(k)/P_{\rm lin}(k)$, where
$P_{\rm rc}(k)$ is measured from the reconstructed linear density
field and $P_{\rm lin}(k)$ is the original linear power spectrum.
In each test, the power spectrum is well recovered on both
large and small scales. However, there is a significant but weak
(about 10\%) dip in the power spectrum ratios. A clear trend is
observed that the wavenumber where the dip appears
(hereafter $k_{\rm d}$) increases with decreasing smoothing scale.

To understand the origin of this discrepancy, we show in the
same figure the average ratio ($R_{\rm F}$) between the likelihood
and prior terms of the Hamiltonian force.
This ratio decreases monotonically with increasing $k$. More
importantly, this ratio is about one at the scale $k_{\rm d}$
(see also the discussion in W13). This strongly suggests that the
dip in the reconstructed linear power spectrum originates from
the competition between the two Hamiltonian force terms.
The Hamiltonian force is the most important quantity that drives the
evolution of $\delta({\bf k})$ in the fictitious system.
At large scales where $R_{\rm F}\gg 1$, the trajectories of
$\delta({\bf k})$ are dominated by the likelihood term so that they
eventually trace well the original linear density field. At small
scales where $R_{\rm F}\ll 1$, on the other hand, the trajectories
of $\delta({\bf k})$ are governed by the prior term, so that the
reconstructed power spectrum matches the original power
spectrum but with totally unconstrained phases.
On scales $R_{\rm F}\sim 1$, where the two terms have approximately
the same importance, the compromise between them
leads to the observed dip. Since the likelihood term increases
with the decrease of the smoothing scale, $R_{\rm s}$, while
the prior term does not, it explains why the dip moves towards
smaller scales as $R_{\rm s}$ decreases.

The deviation observed in Section \ref{sec_asd} also appears around
the scale where $R_{\rm F}=1$, indicative of the same origin.
The question is why a bump appears in the applications to
$N$-body simulations, while a dip is found here. This difference
is clearly due to the inaccuracy of the PM model relative to the
N-body simulation.  An approximate model in general
under-predicts the power spectrum at small scales.
Although such bias is suppressed by smoothing (specified
by $R_{\rm s}$), the reconstructed spectrum is still required
to be enhanced by the HMC to compensate the under-prediction
by the PM model.
The use of a smaller smoothing scale can push the deviation to
a smaller scale. Since $R_{\rm s}>2l_{\rm c}$ is required
(see Section \ref{sec_tpm}), a smaller $R_{\rm s}$ therefore requires
a smaller $l_{\rm c}$ and larger $N_{\rm PM}$.

\newpage

\end{document}